# Heavily Doped Semiconductor Nanocrystal Quantum Dots

David Mocatta[1,3], Guy Cohen[4], Jonathan Schattner[2,3], Oded Millo[2,3†], Eran Rabani[4‡] and Uri Banin[1,3*]

1. Institute of Chemistry, The Hebrew University, Jerusalem 91904, Israel
2. Racah Institute of Physics, The Hebrew University, Jerusalem 91904, Israel
3. The Center for Nanoscience and Nanotechnology, The Hebrew University, Jerusalem 91904, Israel
4. School of Chemistry, The Sackler Faculty of Exact Sciences, Tel Aviv University, Tel Aviv 69978, Israel

To whom correspondence should be addressed: * uri.banin@huji.ac.il, ‡rabani@tau.ac.il, †milode@vms.huji.ac.il

ABSTRACT

Doping of semiconductors by impurity atoms enabled their widespread technological application in micro and opto-electronics. For colloidal semiconductor nanocrystals, an emerging family of materials where size, composition and shape-control offer widely tunable optical and electronic properties, doping has proven elusive. This arises both from the synthetic challenge of how to introduce single impurities and from a lack of fundamental understanding of this heavily doped limit under strong quantum confinement. We develop a method to dope semiconductor nanocrystals with metal impurities providing control of the band gap and Fermi energy. A combination of optical measurements, scanning tunneling spectroscopy and theory revealed the emergence of a confined impurity band and band-tailing. Successful control of doping and its understanding provide *n*- and *p*-doped semiconductor nanocrystals which greatly enhance the potential application of such materials in solar cells, thin-film transistors, and optoelectronic devices.

Doping of bulk semiconductors, the process of intentional introduction of impurity atoms into a crystal discovered back in the 1940s, is a key enabling route for tuning their properties. Its introduction allowed the wide-spread application of semiconductors in electronic and electro-optic components (*1*). Controlling the size and dimensionality of semiconductor structures is an additional powerful way to tune their properties via quantum confinement effects. In this respect, colloidal semiconductor nanocrystals have emerged as a family of materials with size dependent optical and electronic properties. Combined with their capability for wet-chemical processing, this has led to nanocrystal-based light emitting diodes (*2*), solar cells (*3*) and transistor devices (*4*) prepared via facile and scalable bottom-up approaches. Impurity doping in such colloidal nanocrystals still remains an open challenge (*5*). From the synthesis side, the introduction of a few impurity atoms into a nanocrystal which contains only a few hundred atoms may lead to their expulsion to the surface (*6-8*) or compromise the crystal structure. This will inherently create a highly doped nanocrystal under strong quantum confinement, and the electronic and optical properties in such circumstances are still unresolved.

Several strategies have been employed so far for doping nanocrystals. Binding ligands on the nanoparticle surface, which can donate carriers, or electrochemical carrier injection, have been shown to yield *n*-type doping in semiconductor nanocrystal superlattices (*4,9-11*). While of great interest, such remote doping differs from substitutional doping, which has been studied mainly for color center impurities (*12*) and magnetic impurities, notably Mn atoms (*13,14*), providing insight to the challenging chemistry (*15*). Introduction of dopant precursors at specific stages of nanoparticle growth were effective in controlling the impurity location (*16*). More recently, some progress has been made towards producing *n*- type CdSe quantum-dots (QDs) using tin and indium impurities (*17,18*), and *p*-type InP using Cu impurities (*19*).

Here we develop a simple room temperature reaction method for doping semiconductor nanocrystals with metal impurities. By changing the dopant type and concentration, exquisite control of the electronic properties, including the band gap and Fermi energy is achieved. The role of strong quantum confinement leading to localization of the impurity levels, as well as disorder effects leading to band-tailing in small nanocrystals, are studied experimentally and theoretically. Successful controlled doping and its understanding as discovered here, provides *n*- and *p*-doped semiconductor nanocrystals which greatly enhance the potential application of such materials in solar cells, thin-film transistors, and optoelectronic devices.

Adding even a single impurity atom to a semiconductor nanocrystal with a diameter of 4nm, which contains about 1000 atoms, leads to a nominal doping level of $7 \cdot 10^{19} \text{cm}^{-3}$. In a bulk semiconductor this is already within an exceedingly highly doped limit, where metallic ('degenerate') behavior is expected (*20*). Doping at this level in bulk semiconductors leads to several effects summarized in Figure 1A. First, at high doping levels the impurities interact with each other and an impurity sub-band emerges near the edge of the respective band (conduction or valence for *n*- or *p*-type, respectively). Often, tail states (Urbach tail) also develop due to distortions in the crystal structure (*21*). Effectively, the band gap, $E_g$ is narrowed. This may be probed by optical means where for highly doped *n*-type semiconductor (left frame in Figure 1A), the absorption is blue shifted due to conduction band filling by the donated electrons (Moss-Burstein effect) (*20*), and the emission emanating from the bottom of the conduction band is red-shifted. For high *p*-type doping, both absorption and emission are typically seen to be red-shifted due to the complexity of the valence band dominated by the band-tailing effect (*20*).

A dramatically different situation arises for impurity doping of nanocrystals due to the discrete nature of the quantum confined states (Figure 1B, for *n*-type doping). In this case the addition of a single dopant (left frame) significantly alters the density of states (DOS) due to the introduction of the impurity levels, a situation which is not expected for prior remote surface doping strategies. This has been described by a hydrogenic model under spherical confinement, leading to *S*- and *P*-like impurity states denoted by $E_d^{1S}$ and $E_d^{2P}$ that are several tens of meVs ($\Delta$) below the corresponding dot levels, effectively doubling the DOS near those energies (*22-24*). Even more intriguing is the case introduced here, of multiple impurities in a single dot, which are inherently interacting due to the small volume and experience the effect of the confining potential (Figure 1B, right frame). Under these conditions, the nature of delocalization and interaction of the impurity charge carriers may be greatly modified as compared to the bulk case. In addition, multiple impurities in a small confined nanocrystal can enhance disorder effects, altering the electronic structure via a quantum-confined Urbach tail mechanism.

To dope InAs nanocrystals with different impurity atoms, we modify a reaction used for gold growth onto semiconductor nanoparticles (*25*). In a typical metal atom doping synthesis, the metal salt, $CuCl_2$, $AgNO_3$, $AgCl$, or $AuCl_3$, is dissolved in a toluene solution with appropriate surfactant and gradually added to a solution of InAs QDs in toluene at room temperature (*26*).

Figure 2A shows a transmission electron microscope (TEM) image of Ag doped QDs (Figure S8). At the impurity levels reported below, it was not possible to identify the presence of metal regions, indicating that the impurity atoms are dispersed. We note in passing, that this was not the case for very high metal atom concentrations (~3000 atoms per QD in the reaction solution), where TEM analysis clearly showed phase separation between InAs and impurity metal regions (*25*). Further support for the dispersion of impurities is provided by X-ray diffraction (Figure S10), where no fingerprints of metal domains were detected while the InAs crystal structure is generally maintained. Some broadening of the peaks is seen, ascribed to a small degree of structural disorder. X-ray photoelectron spectroscopy (XPS) measurements of these samples were also performed indicating the presence of Ag, Au, and Cu in the respective samples (Figures S11 and S12). This suggests successful addition of these atoms to the InAs QDs, consistent with our previous report of Au growth on InAs QDs, which exhibited the room temperature solid state diffusion of Au into the QD (*25*). Indeed, extrapolating the diffusion parameter values to room temperature gives a diffusion length scale of $~10^4$ nm/24h, far greater than the QD diameter, and large values are also extrapolated for Ag and Cu (Table S1).

Figure 2B shows the absorption and emission (inset) spectra of undoped and doped InAs QDs. The addition of Ag atoms results in a red-shift of both the first exciton absorption and the emission peaks. The addition of Cu results in a blue-shift of the first exciton absorption peak whereas the emission is not shifted. Addition of Au at similar concentrations does not significantly alter the observed optical gap neither in absorption nor in emission. The addition of any of these impurity atoms results in the gradual quenching of the emission from the QDs, yet the three impurities lead to qualitatively different effects on the optical spectra and hence on the electronic properties of the doped QDs. The effect of varying amounts of impurities on the first absorption peak and on the emission is shown in Figure 2C-E for InAs QDs with different diameters (see Figures S1-2 for individual spectra). The amount of impurities on the QDs was estimated by the analytical method of inductively coupled plasma atomic emission spectroscopy (ICP-AES, Figures S3-S7).

A first possible source of spectral shifts in such quantum confined particles may be related to size changes upon doping, but this was excluded by detailed sizing analysis (Figure S9). An alternative source of the spectral shifts can be associated with electronic doping by the impurities. In Figure 3 we show tunneling spectra measured by a scanning tunneling microscope (STM) at T=4.2K for undoped, Au-doped, Cu-doped and Ag-doped InAs QDs 4.2nm in diameter (see Figures S13-S15 for additional spectra drawn for extended voltage scale). Starting from the reference case of the undoped QD shown in the lower panel, the dI/dV curves, which are proportional to the DOS, match earlier studies of InAs QDs *(27)*. The gap region is clearly identified, while on the positive bias side a doublet of peaks associated with tunneling through the doubly degenerate $1S_e$ state is seen at the onset of the current, followed by higher order multiplet at higher bias corresponding to tunneling through the $1P_e$ conduction band state. A more complex peak structure is seen on the negative bias side, resulting from tunneling through the closely spaced and intricate valence band states.

Several changes are seen upon doping the QDs. Starting with the case of Au, the gap is similar to the undoped NC, consistent with the optical measurements. However, the features in the scanning tunneling spectroscopy (STS) spectra are washed out, suggesting that indeed Au has entered the nanocrystal, perturbing the pristine level structure. More significant changes are seen for both the Cu and Ag cases, presented in the upper panel. Significant band-tailing into the gap and emergence of in-gap states in regions covering nearly 40% of the gap region are observed. In particular, in the Cu case, a shoulder on a tail-state structure is seen at bias values just below the $1S_e$ doublet (which is remarkably preserved). Additionally, the doublet is superimposed on a notable rising background that increases to the region of the 1P$_e$ peaks that are not well resolved (Figure S13). For the Ag-doped QDs, there is significant broadening and merging of features on the positive bias side, and on the negative bias side a background signal develops.

A clear result of doping in bulk semiconductors is the shift of the Fermi level, which for *n*-type doping is close to the conduction band, and conversely, shifts to a lower energy close to the valence band, for *p*-type impurities. Remarkably, such shifts are clearly identified in the STS of the Cu and Ag doped QDs measured by the positions of the band edges relative to zero bias. While the Fermi energy for the undoped case, as well as the Au doped case, is nearly centered, in the Cu-case the onset of the conduction band states nearly merges with the Fermi energy consistent with *n*-type doping. In contrast, for the Ag case, the Fermi level is much closer to the onset of the valence band states, signifying *p*-type doping in this case.

Chemical considerations for the doping of InAs with the different metal atom impurities can help to understand these observations (Figure S17). Cu can have a formal oxidation state of either $Cu^{2+/1+}$. Moreover, its ionic radius is the smallest of the three impurities and therefore may be accommodated in interstitial sites within the InAs lattice (*28*). In such a case, one can expect that the Cu will partly donate its valence electrons to the QD leading to *n*-type doping, consistent with the shift in the Fermi energy observed by STS. The incorporation of multiple impurities is expected to lead to the development of closely spaced impurity states, akin to the impurity band formed in the bulk (*20*). This band forms asymmetrically due to the disordered arrangement of the impurities in the QD, surpassing the energy of the 1S$_e$ NC state. The observed rising background in the STS curve signifies the presence of such an impurity band. This is a direct indication of the substantial modification of the DOS induced by the impurities in small QDs, corresponding to very highly doped behavior in the bulk. Revisiting the observed blue shift in the absorption, this is in line with the filling of the conduction and asymmetric impurity-band levels in heavily *n*-type doped QD, leading to a Moss-Burstein blue-shift in the absorption spectrum and minor shifts in the emission (Figure 2).

Ag has a large radius, and is considered to be a substitutional impurity in III-V semiconductors (*29*). The replacement of an In atom, which possesses three valence electrons, with a Ag atom, which has only one valence electron, leads to an electron deficiency in the bonding orbitals causing *p*-type doping. This is reflected in the shift of the Fermi level, as seen in the STS data (Figure 3, red trace). In this case, the rising background in the spectrum at negative bias, indicates the formation of an impurity band near the valence band. Since Ag has the largest ionic radius of the three impurities, it distorts the crystal structure most signficantly; this results in band-tailing analogous to the Urbach tail known for highly doped bulk semiconductors, and leading to the red shifts observed in the absorption onset as well as the emission (Figure 2). Au may adopt a +3 valence state, which makes it isovalent with In and hence doping is not expected to lead to the introduction of charge carriers. Moreover, its size is comparable to that of the In (Table S1), allowing for substitutional doping without significant lattice distortions. These features of Au are consistent with the absence of significant shifts in absorption, emission and Fermi energy, as observed in both optical and tunneling spectra.

For deeper insights we have considered modeling the effects of electronic doping and structural disorder on the electronic properties of strongly confined impurity dopants in QDs. Starting with electronic effects, two models representing two limiting cases of doping were developed (SOM for more details). The starting point for both models is based on the hydrogenic-like impurity model (*22-24*). In the first limit ("Z-model", Figure 1B) we assume weak localization of the impurity states represented as a single central multivalent impurity. In the other limit impurity electrons are sufficiently localized, to an extent much greater than in bulk (*8*), such that a single-electron tight binding (TB) treatment where each impurity contributes a single electron is applicable. In Figure 4A, we show the impurity DOS for *n*-type doping of two nanocrystal sizes for the TB model. For small number of impurities (e.g. N=5), a peak in the DOS is observed below the nanocrystal level ($1S_e$). As the number of impurities increases, an impurity band develops asymmetrically around the nanocrystal level, pushing the Fermi energy towards the conduction band. The emerging impurity band develops a tail which extends into the gap consistent with the STS data for the Cu case.

In Figure 4B we show TB estimations for the shifts in the absorption spectrum as a function of the QD radius for various doping levels. The single impurity limit is the solution of a finite barrier confined hydrogenic impurity, and nearly scales as $R^{-1}$. In this case, for both types of doping, a red shift is calculated. Although the confined energy levels of the impurity and the QD depend strongly on the effective mass ($m^*_{e,h}$), the spectral shifts, which are given by the differences between the impurity and electron/hole levels, only depend weakly on $m^*_{e,h}$ (Figure 4A). To lowest order in the Coulomb interaction, these differences are given by the expectation value of the electrostatic energy and depend weakly on $m^*_{e,h}$. As the number of impurity atoms increases, we find a non-monotonic dependence of the spectral shifts with nanocrystal size. This reflects the transition of the impurity DOS from the gap region into the conduction band.

The onset of the blue shift and its magnitude depend on the model parameters (Figure 4C). At low impurity concentrations, a red shift is always observed, consistent with the single hydrogenic impurity limit. A turnover from negative to positive shift is observed, with increasing impurity density, depending on the degree of impurity carrier localization, determined by the ratio of the range of impurity interactions $a$ and QD size $R$. The regime at which a blue shift is observed sets in earlier as $a/R$ increases. DFT calculations (*30*) for single impurities in nanocrystals suggest $a/R \approx 1/3$, implying a narrow regime of red shift before a blue shift sets in. This is qualitatively consistent with the experimental observations shown in Figure 4D for Cu, where a red shift is absent, suggesting that $a < R$. Moreover, the size dependence of the blue shift, which decreases with increasing $R$, is consistent with the *R*-dependence shown in Figure 4C for the theory. The theory predicts that the shift will eventually reach a plateau, which is not observed experimentally.

The asymmetry in the shift (Figure 4C) is directly correlated with the magnitude of the hopping term, $\gamma$. The case shown in Figure 4C is for $\gamma = 5\varepsilon$ ($\varepsilon$ is the single impurity energy taken from the hydrogenic model), where the blue shift is significantly larger. When $\gamma = \varepsilon$ one observes a symmetric shift around the nanocrystal level (Figure S22). Also shown in Figure 4C are the results of the *Z*-model, yielding an opposite behavior (inconsistent with current experiments), where a blue shift is observed for low doping levels, turning to a red shift as the doping level increases. These fundamental differences between the *Z*- and TB-models suggest that confinement leads to localization, consistent with previous theoretical work on single dopants (*8*) and the neutral impurity oxidation states observed in the XPS measurements.

A second important mechanism of electronic level modification in the regime of highly doped semiconductors relates to distortions of the crystal structure by the dopants leading to band-tailing and a red shift. To address the role of band-tailing we employed an atomistic treatment of the electronic structure of the nanocrystal, where disorder was introduced by randomly displacing In or As atoms (*26*). Band-tailing occurs mainly for the valence band edge as evident from the DOS shown in Figure 4E for different impurity numbers. This results from the heavier effective mass and the denser level structure of the valence band, compared to the light effective mass of the highly delocalized electron. From the DOS we estimate the shifts in the band gaps induced by disorder as shown in Figure 4F. Tailing occurs for distortions to the In atoms, As atoms, both atoms or only surface atoms. The band-gap decreases for higher dopant concentrations. Note that the As distortions lead to larger shifts, consistent with its larger ionic radii.

In Figure 4G and H we show the calculated and measured shifts (for Ag impurities) for several nanocrystal sizes as a function of the dopant concentration, respectively. The striking similarities between the theory and experiments suggest that the spectral shifts observed in Ag, unlike the case of Cu, are dominated by band-tailing. Both theory and experiments show a rapid increase in the magnitude of the shift, reaching a plateau at high dopant concentrations. The increase in the shift depends on the level of disorder induced by each dopant. Furthermore, the shifts decrease with increasing NC size in both cases. We note that in Cu, while there maybe also a contribution of band-tailing, the band-filling effect discussed above dominates, leading to the blue shift in that case.

Doping semiconductor nanocrystals with metal impurities provides further means to control their optical and electronic properties. We developed a simple solution based method to dope colloidal nanocrystals, and demonstrated *n* and *p*-type doped InAs nanocrystals by introducing Cu and Ag impurities, respectively. Cu doped particles showed a blue shift in the absorption and no shift in emission, in line with partial filling of conduction band states with electrons from interstitial Cu impurities. Correspondingly, the STS measurements show a shift in the Fermi energy to near the conduction band edge, and the development of a confined impurity band. This behavior was rationalized by a TB model, and its size dependence indicates that the confinement leads to localization of the impurity states in small nanocrystals. Conversely, STS measurements of Ag-doped nanocrystals led to a shift of the Fermi energy towards the valence band, proving *p*-type doping. This was accompanied by a red shift of both absorption and emission peaks, attributed to band-tailing effects analogous to the Urbach tail, in line with Ag adopting substitutional sites. The magnitude and size dependence of the band narrowing agrees with atomistic electronic structure calculations incorporating structural disorder induced by the dopants. Interestingly, introduction of Au impurities as a substitutional dopant, although broadening the STS spectra, maintains the position of the Fermi level and does not lead to spectral shifts, in line with its isovalent nature with In. The controlled ability to synthesize *n*- and *p*-type doped nanocrystals, along with the deep understanding of the highly doped impurity regime in colloidal QDs, opens avenues for diverse electronic and optoelectronic devices.

**Figures**

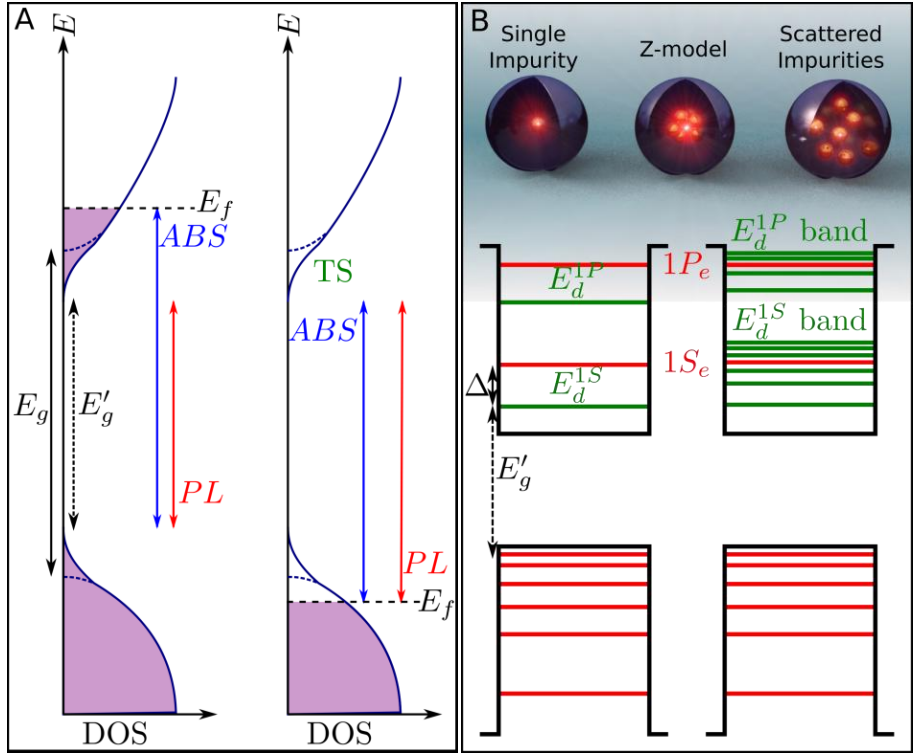

Figure 1. Diagrams describing the effects of heavy doping in bulk and nanocrystal semiconductors. (A) A scheme of the different influences of doping a bulk semiconductor for *n*-type (left) and *p*-type (right) dopants. ABS – absorption onset, PL – photoluminescence onset, TS – tail states, $E_f$ - Fermi energy, $E_g$ - tailed band gap, $E_g'$ - unperturbed band gap. The purple shading shows state filling up to the Fermi energy. (B) A sketch for *n*-doped nanocrystals QD with confined energy levels, red and green line correspond to the QD and impurity levels, respectively. Left: The level diagram for a single impurity effective mass model, where $E_g'$ is the quasi-particle gap in the doped QD, $1S_e, 1P_e$ are the QD electron levels and $E_d^{1S}, E_d^{1P}$ are the impurity levels shifted below the corresponding QD levels by $\Delta$. Right: Impurity levels develop into impurity bands as the number of impurities increases. Upper panel shows a sketch of the different impurity models.

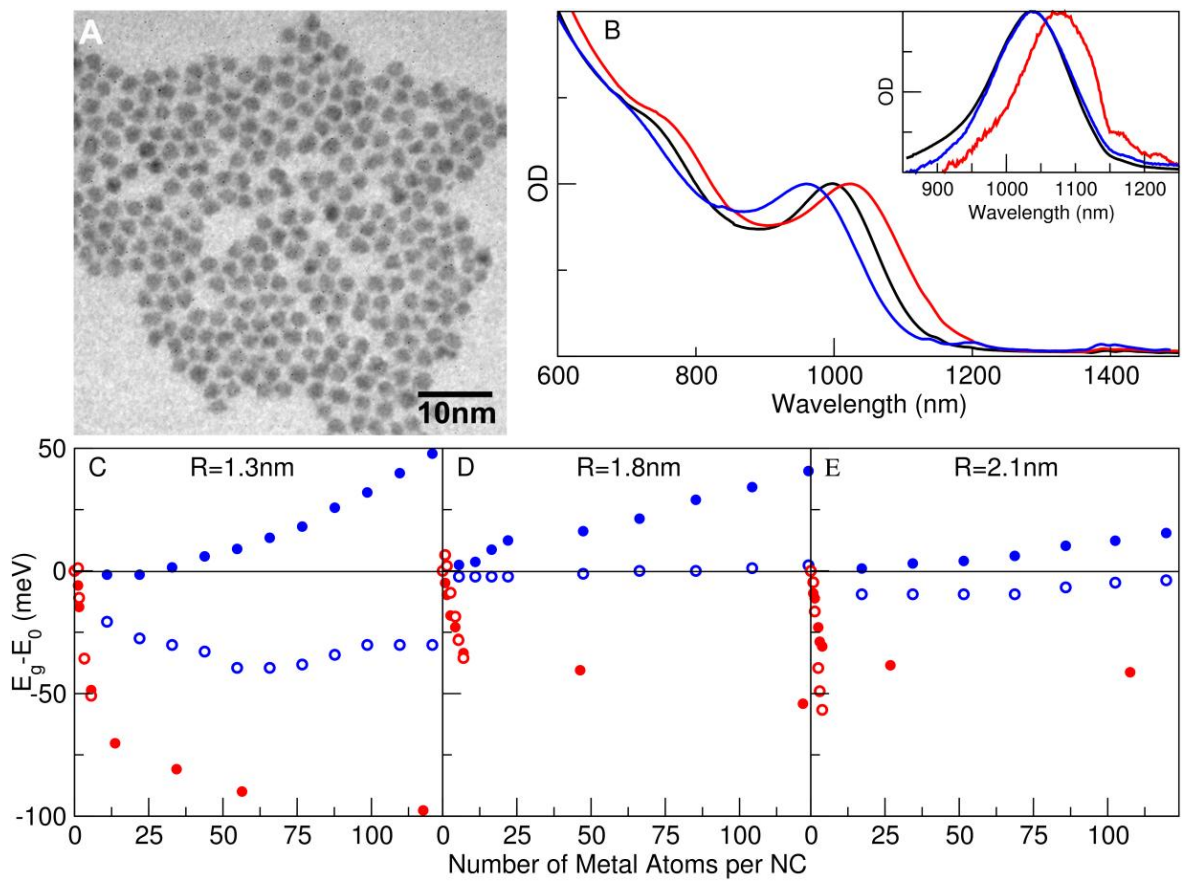

Figure 2. Optical properties of doped InAs NCs. (A) A TEM image of Ag doped 3.3 nm InAs NCs. (B) The normalized absorption spectra of three 3.3nm diameter InAs NC samples. Two of the samples had Cu (Blue) and Ag (Red) solutions added to them resulting in metal:NC solution ratios of 540 and 264 respectively. These amounts correspond to 73 Cu atoms and 9 Ag atoms per NC. The third sample had a control solution (without metal salt) added to it (Black). The inset shows the normalized emission spectra of these samples. (c-e) The energetic shift of the first exciton peak (closed symbols) and the emission energy (open symbols) against the number of impurity atoms per QD for InAs NCs with radii of 1.3nm (C), 1.8nm (D) and 2.1nm (E). Red symbols correspond to Ag doping, blue symbols to Cu.

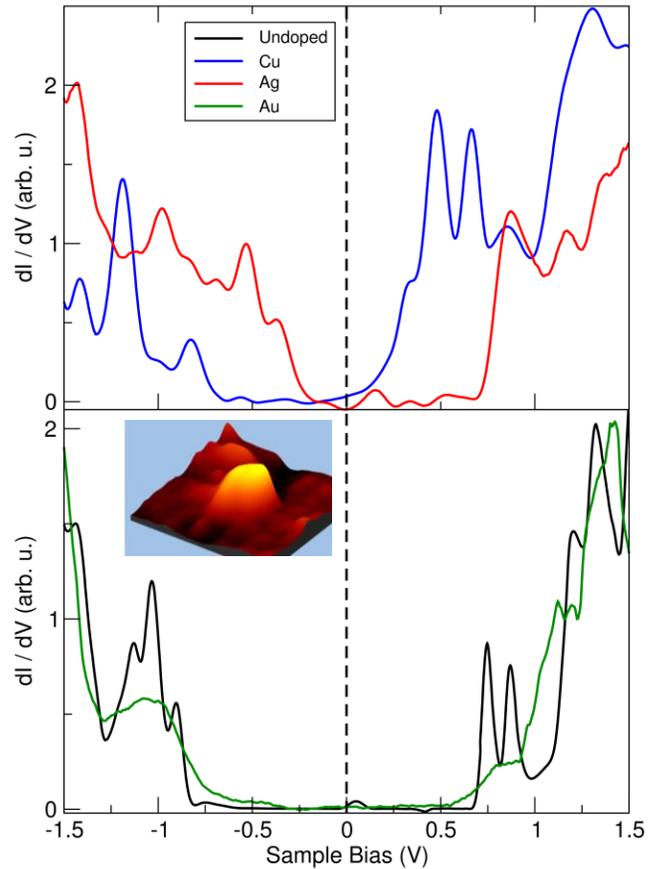

Figure 3. The effect of doping on the STM tunneling spectra. Four dI/dV *vs.* V tunneling spectra at 4.2K, of undoped (black trace), Au-doped (green trace), Cu-doped (blue trace) and Ag-doped (red trace) InAs nanocrystals, nominally 4 nm in diameter. The doped QDs were taken from samples that had Ag, Cu and Au atom to NC ratios corresponding to 15, 160 and 77, respectively. The vertical (V=0) dashed line is a guide to the eye, highlighting the relative shifts of the band edges in the doped samples in manner typical of *p*-doped and *n*-doped semiconductors for the Ag-doped and Cu-doped nanocrystals, respectively. The inset shows an STM image of a single (Ag-doped) QD on which STS data were measured.

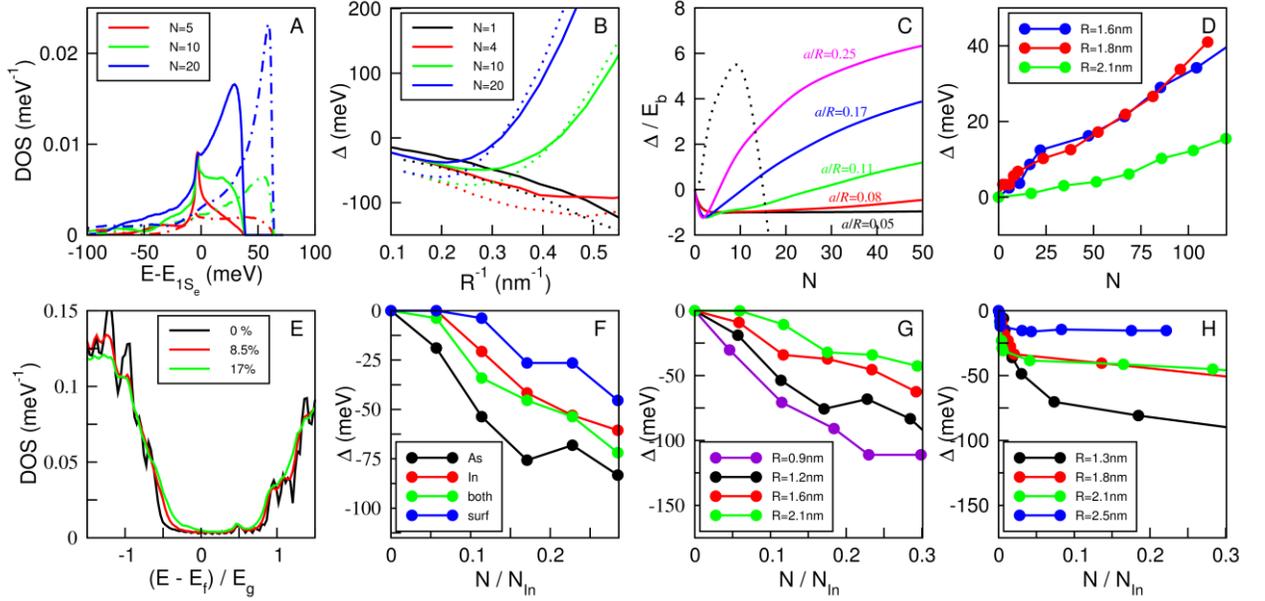

Figure 4. The effect of doping on the electronic structure of nanocrystals. (A-D) Impurity band formation, (A) The impurity DOS from the TB model ($a = 0.4$nm) for different levels of doping for $R = 2$nm (solid curves) and $R = 1.3$nm (dotted-dashed curves). (B) The spectral shifts as a function of the inverse NC size for *n*-type (solid curves) and *p*-type (dotted curves) doping for different numbers of impurities. The InAs levels were calculated from an effective mass model. The impurity levels were calculated from the TB model with $a = 0.4$nm. (C) Spectral shifts for *n*-type doping, as a function of the number of impurities $N$ in units of the binding energy, $E_b$, of a single hydrogenic impurity, for several levels of localization ($a/R$). Dashed curve is the result of the Z-model. (D) Experimental spectral shifts for Cu doping as a function of $N$ for several NC sizes. (E-H) Band tailing in doped nanocrystals. (E) DOS of an $In_{140}As_{141}$ NC for three levels of doping. (F) Shifts in gap as a function of $N$ relative to the number of In atoms ($N_{In}$) in the NC for different ways of introducing structural disorder. (G) Calculated shifts induced by band-tailing for several NC sizes. (H) Measured spectral shifts for Ag doped NCs for several NC sizes. We used a normal distribution to induce disorder with a width that is 10% the In-As bond length (*26*).

**Acknowledgment**

The authors acknowledge the help of the staff of the Unit for Nanocharacterization of the Center for Nanoscience and Nanotechnology in the Hebrew University, Jerusalem, headed by Dr. I. Popov, and Dr. Vitaly Gutkin for help in the XPS studies. This research was supported by a grant from the ERC project DCENSY (UB) and by the Israel Science Foundation (O.M.). O.M. acknowledges support from the Harry de Jur Chair in Applied Science. U.B. thanks the Alfred and Erica Larisch Memorial Chair. G.C. is grateful to the Azrieli foundation for the award of an Azrieli fellowship. D.M. acknowledges support from the Centre for Scientific Absorption, the Ministry of Absorption, the state of Israel.

Supplementary Online Material for

# Heavily Doped Semiconductor Nanocrystal Quantum Dots

David Mocatta[1,3], Guy Cohen[4], Jonathan Schattner[2,3], Oded Millo[2,3†], Eran Rabani[4‡]

and Uri Banin[1,3*]

1. Institute of Chemistry, The Hebrew University, Jerusalem 91904, Israel

2. Racah Institute of Physics, The Hebrew University, Jerusalem 91904, Israel

3. The Center for Nanoscience and Nanotechnology, The Hebrew University, Jerusalem 91904, Israel

4. School of Chemistry, The Sackler Faculty of Exact Sciences, Tel Aviv University, Tel Aviv 69978, Israel

To whom correspondence should be addressed: * uri.banin@huji.ac.il, ‡ rabani@tau.ac.il, † milode@vms.huji.ac.il

## Materials

In(III)Cl$_3$ (99.999+%), tris(trimethylsilyl) arsenide (TMS$_3$As), trioctylphosphine (TOP, 90%; purified by vacuum distillation and kept in the glovebox), AuCl$_3$ (99%), AgCl (99+%), AgNO$_3$ (99+%), CuCl$_2$ (99.999%), dodecylamine (DDA, 98%), didodecyldimethylammonium bromide (DDAB, 98%), toluene (99.8% anhydrous), methanol (99.8% anhydrous). All chemicals were purchased from Sigma Aldrich except for (TMS$_3$As) which was synthesized as detailed in the literature (*1*).

## Methods

### InAs Nanocrystal Synthesis

The synthesis of InAs nanocrystals (NCs) is based on the literature procedure (*2*) and is carried out under an inert atmosphere using standard Schlenck techniques. In a typical synthesis a mixture of indium and arsenic precursors is prepared by adding 0.3g (1mmol) of (TMS$_3$As) to 1.7g of a 1.4M InCl$_3$ TOP solution (2mmol in total). 1ml of this solution is injected into a three neck-flask containing 2ml of TOP at 300°C under vigorous stirring. The temperature is then reduced to 260°C and further precursor solution is added in order to achieve particle growth. The growth was monitored by taking the absorption spectra of aliquots extracted from the reaction solution. Upon reaching the desired size, the reaction mixture was allowed to cool to room temperature and was transferred into the glovebox. Anhydrous toluene was added to the reaction solution, and the nanocrystals were precipitated by adding anhydrous methanol. The size distribution of the nanocrystals in a typical reaction was on the order of 10%. This was improved using size selective precipitation with toluene and methanol as the solvent and anti-solvent, respectively.

### Metal-atom Doping Method

In a typical reaction a metal solution is prepared by dissolving 10mg of the metal salt ($CuCl_2$, $AgNO_3$, $AgCl$ or $AuCl_3$), 80mg DDAB and 120mg of DDA in 10ml of toluene. The Cu and Ag solutions prepared in this manner are respectively blue, colorless and yellow. The metal solution is then added drop-wise to a stirred 2ml toluene solution of InAs NCs. After 15 minutes the absorption and emission of the solutions are measured. The Cu and Au samples are precipitated with methanol whilst the Ag sample is precipitated with acetone. The entire metal treatment procedure is carried out under inert conditions. The ratio of metal atoms to NCs in solution is estimated from the literature values of InAs NC absorption cross-sections (*3*).

### Instrument Specifications & Sample Preparation

UV-Visible- Near IR Spectroscopy was performed on a JASCO V-570 spectrometer using glass cuvettes. The photoluminescence (PL) measurement with an Ocean-optics NIR-512 spectrophotometer using 633nm HeNe laser excitation. Transmission Electron Microscopy was performed on Tecnai T12 G2 Spirit at 120 keV. TEM grids were prepared by evaporating one drop of the metal treated NC solution onto a standard carbon covered grid. This was followed by washing the grid with several drops of methanol for Cu and Au samples and acetone for Ag samples.

Powder X-ray Diffraction (XRD) patterns were obtained using Cu Kα photons from a Phillips PW1830/40 diffractometer operated at 40 kV and 30 mA. Samples were deposited as a thin layer on a low-background scattering quartz substrate. The data presented has undergone a background subtraction.

X-ray photoelectron spectroscopy (XPS) was carried out with a Kratos Analytical Axis Ulta XPS instrument. Data were obtained with Al KR radiation (1486.6eV). Survey spectra were collected with 160eV pass energy detection. The measurements were performed on nanocrystal films of monolayer thickness deposited on a Si wafer.

ICP-AES measurements were carried out using a Perkin-Elmer Optima 3000. Emission from In ions was measured at 230.6nm and 325.6nm, Emission from As ions was measured at 189.0nm 193.7nm. Cu, Ag and Au emission was measured at 324.8nm, 328.1nm and 208.2nm respectively. The sample preparation procedure is described later.

### Detailed Spectra of Samples During the Doping Procedure

The treatment of the InAs NCs with the Cu solution leads to an increase in the first exciton energy (see Figure S1 panels a-c) while the emission energy is unchanged (see Figure S1 panels d-f).

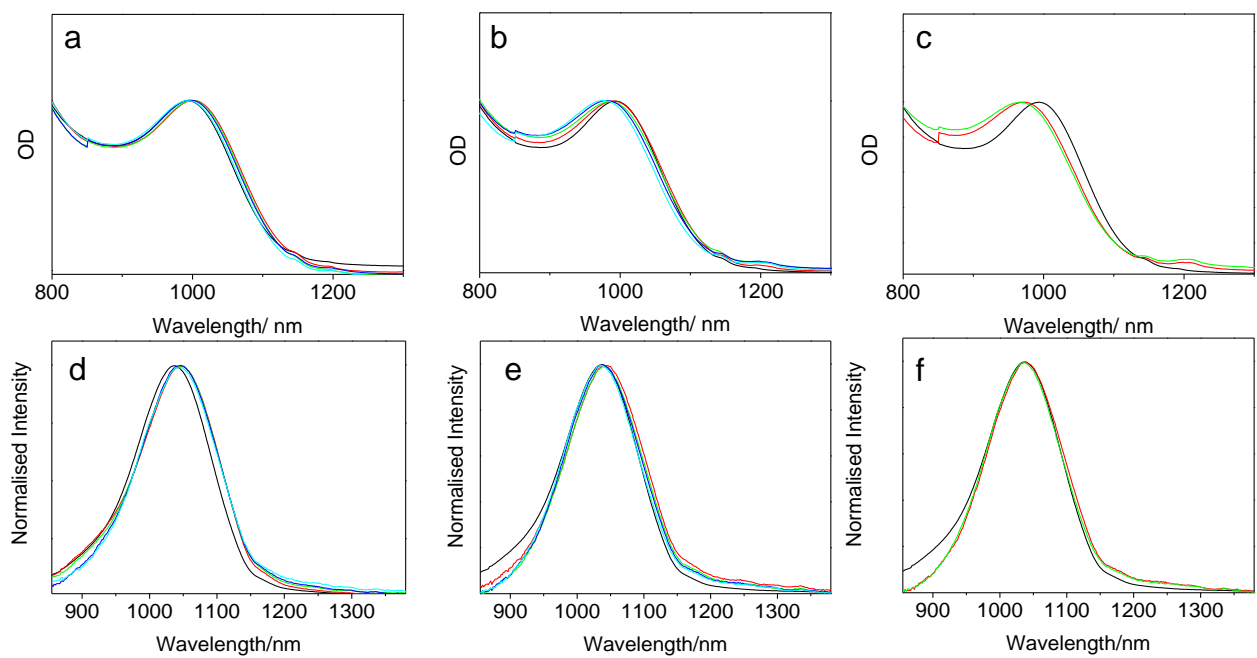

Figure S1. Sample spectra for Cu doping. The normalized absorption (a-c) and emission spectra (d-f) for 3.3nm diameter InAs NCs to which increasing amounts of Cu solution were added. The Cu:NC solution ratios are as follows: panel (a & d) 0 (Black), 85 (Red), 170 (Green), 260 (Blue), 340 (Cyan); panel (b & e) are 0 (Black), 425 (Red), 510 (Green), 600 (Blue), 680 (Cyan); panel (c & f) are 0 (Black), 770 (Green), 850 (Green).

Ag treatment leads to a decrease both in the first exciton energy, see Figure S2 panels a and b and the emission energy (see Figure S2) panels c and d. All the metal treatments result in some quenching of the emission. This is totally quenched after precipitation.

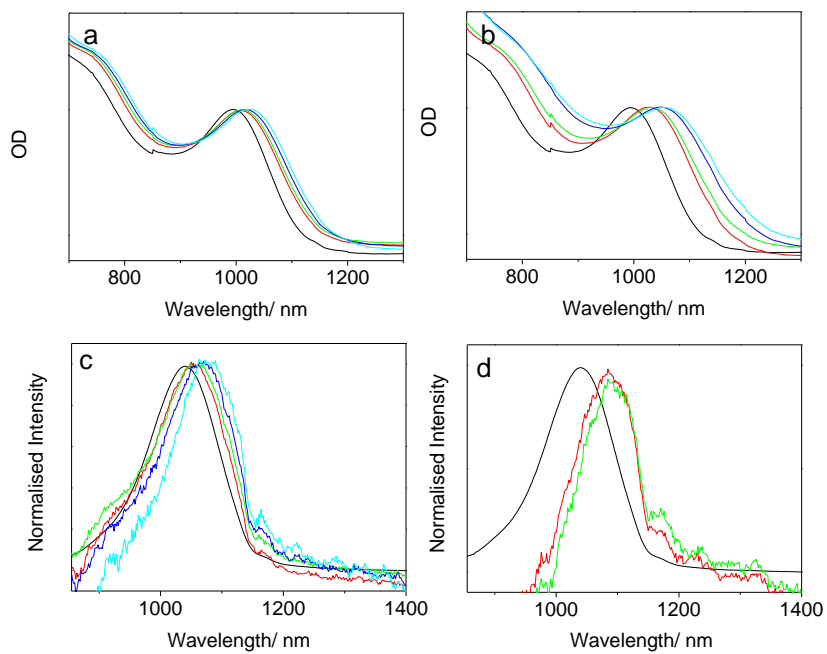

Figure S2. Sample spectra for Ag doping. The normalized absorption (a & b) and emission spectra (c & d) for 3.3nm diameter InAs NCs to which increasing amounts of Ag solution were added. The Ag:NC solution ratios in are as follows: panel (a & c) 0 (Black), 50 (Red), 90 (Green), 170 (Blue), 260 (Cyan); panel (b & d) are 0 (Black), 332 (Red), 440 (Green), 850 (Blue), 1000 (Cyan).

**Determination of Metal Impurity Concentration by ICP-AES (Inductively Coupled Plasma Atomic Emission Spectroscopy) Measurements**

In order to estimate the number of metal atoms per NC, metal treatment was carried out as described above on a range of NC sizes and at different metal concentrations. The NCs were then precipitated, dissolved using concentrated $HNO_3$ and diluted with triply distilled water. The relative quantities of In, As, Cu, Ag and Au in these solutions was measured by ICP-AES. The amount of Cu/Ag/Au per NC was extrapolated using the ratio of In and As to Cu/Ag/Au atoms together with the estimated number of In and As atoms per NC estimated from the NC size. The results are shown for Cu and Ag in Figure S3 and Figure S5. These graphs represent the number of metal atoms per NC extrapolated from the ICP measurements against the metal:NC solution ratio.

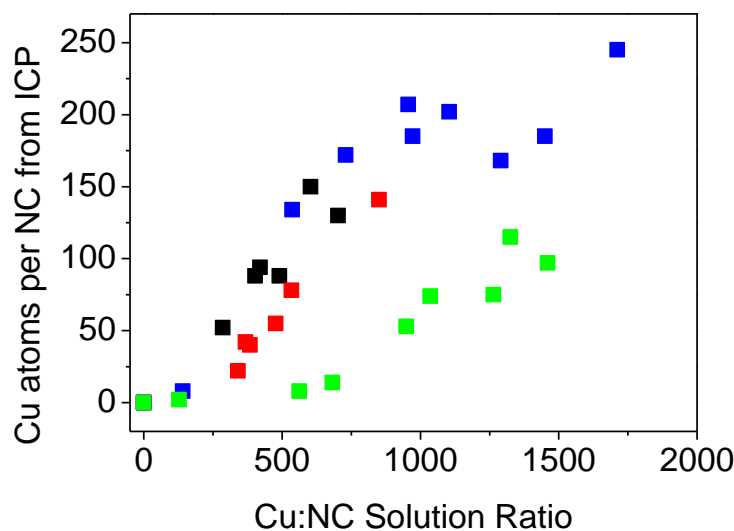

Figure S3. The number of Cu atoms per NC. Extracted values from ICP measurements against the Cu:NC solution ratio for NCs with diameters of 2.7nm (Black), 3.3nm (Red), 3.6nm (Green) and 4.1nm (Blue).

Figure S3 shows the dependence of the number of Cu atoms per NC on the Cu:NC solution ratio. For each size the results were fitted and the interpolated relationship was used to calibrate the number of Cu atoms per NC reported here. Figure S4 is an example of a fit for the data of 3.6nm diameter NCs. In this case the data was separated into two regions which had linear fits applied.

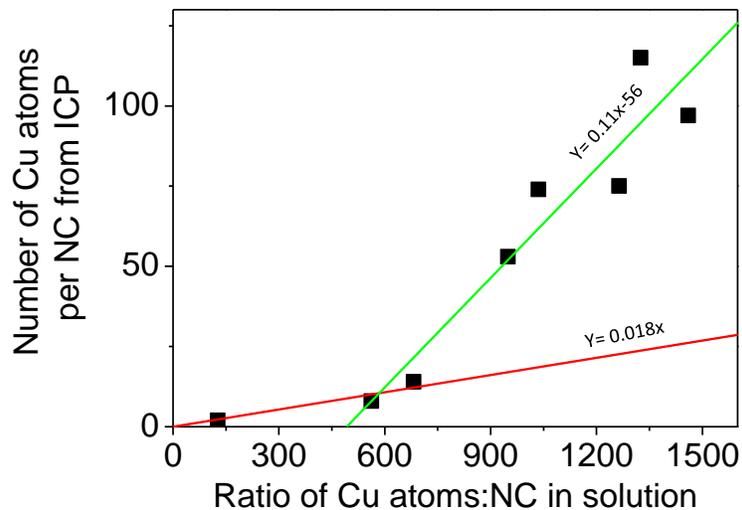

Figure S4. Calibration of the relationship of Cu per NC. The number of Cu atoms per NC extracted from ICP measurements versus Cu:NC solution ratio for 3.6nm diameter NCs. The data was separated into two regions to which linear fits were applied for interpolation (Red and Green).

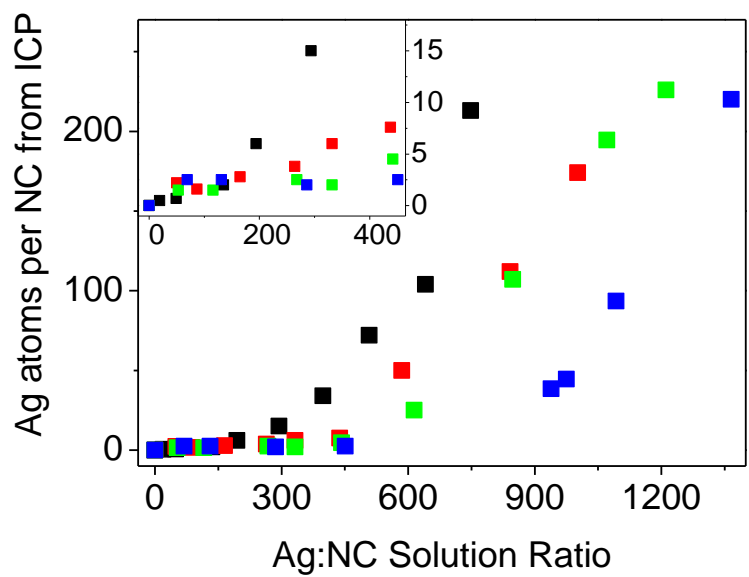

Figure S5. The number of Ag atoms per NC. The number of Ag atoms per NC extracted from ICP measurements against the Ag:NC solution ratio for NCs with diameters of 2.7nm (Black), 3.3nm (Red), 4.1nm (Green) and 5.0nm (Blue). The inset is an enlarged representation of the region where the Ag:NC solution ratio is less than 450.

The number of Ag atoms per NC on the Ag:NC solution ratio showed two regimes as can be seen in Figure S5. For the purpose of calibration these regions were fitted separately an example of which is shown in Figure S6. Similar fits were calculated for all particle sizes and the results were used to calibrate the number of Ag per NC reported here. Similar analysis was undertaken for Au.

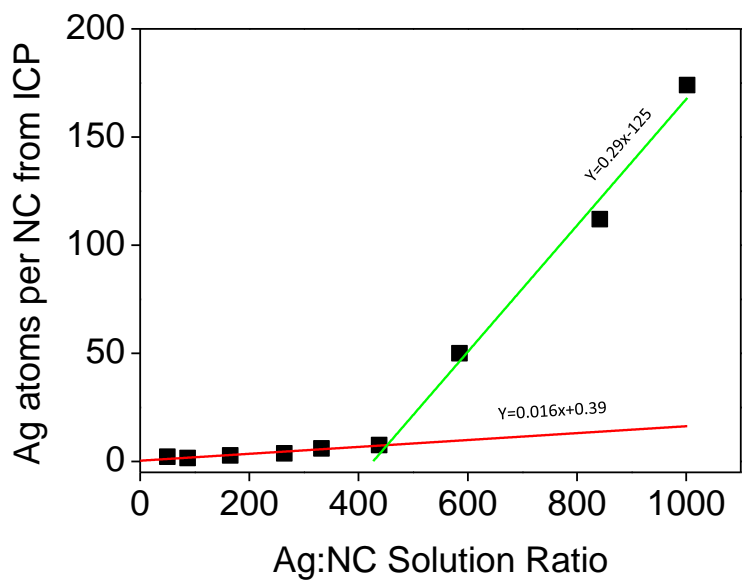

Figure S6. Calibration of the relationship of Ag per NC. The number of Ag atoms per NC extrapolated from ICP measurements against the Ag:NC solution ratio for 3.3nm diameter NCs. The data was separated into two regions to which linear fits were applied (Red and Green).

## Summary of Optical and Chemical Data for Ag Doping

Figure S7 shows the Ag:NC solution ratio (top frame), and shift of the first excitonic absorption peak (bottom frame) against the number of Ag atoms per NC for InAs NC, extracted from ICP-AES. At Ag:NC solution ratios below 500 relatively small numbers of Ag atoms are present on the NC, ~<10. At these low levels the shift dependency exhibits a sharp nearly linear response. For Ag:NC solution ratios above 500 the relationship changes and much larger numbers of Ag atoms are found on the NCs. The sudden change in the relationship between the number of Ag atoms per NC and the Ag:NC solution ratio suggests that the Ag growth mode has changed from being dispersed within the lattice to growing on the surface of the particle. This change corresponds to the onset of saturation in the shift magnitude and so indicates that the saturation of the shift is due to the change in the growth mode to surface growth.

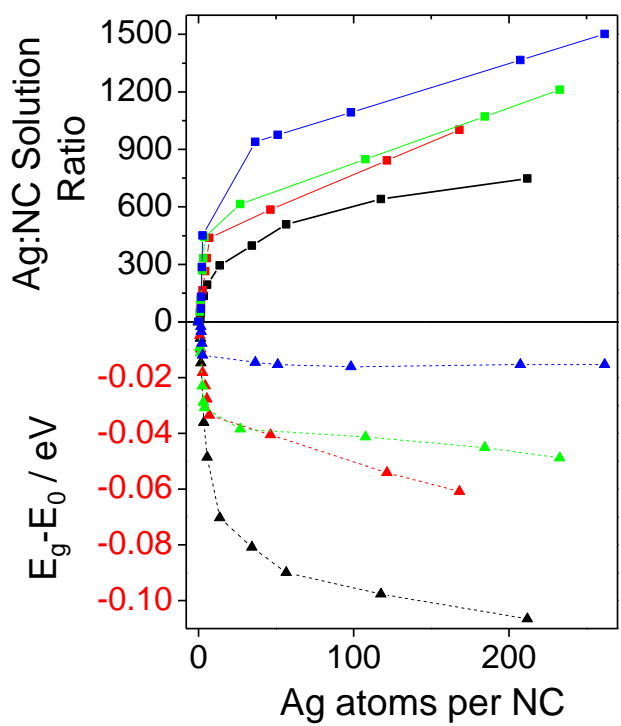

Figure S7. The number of Ag atoms per NC and shift data. The Ag:NC solution ratio (upper-left axis, square symbols, solid line) and first exciton shift (lower-left axis, triangular symbols, dashed line) against the number of Ag atoms per NC for InAs NCs of various radii 1.3nm (Black), 1.7nm (Green), 2.1nm (Red) and 2.5nm (Blue).

**Additional Information and Experimental Details for Characterization**

**TEM Sizing Characterization**

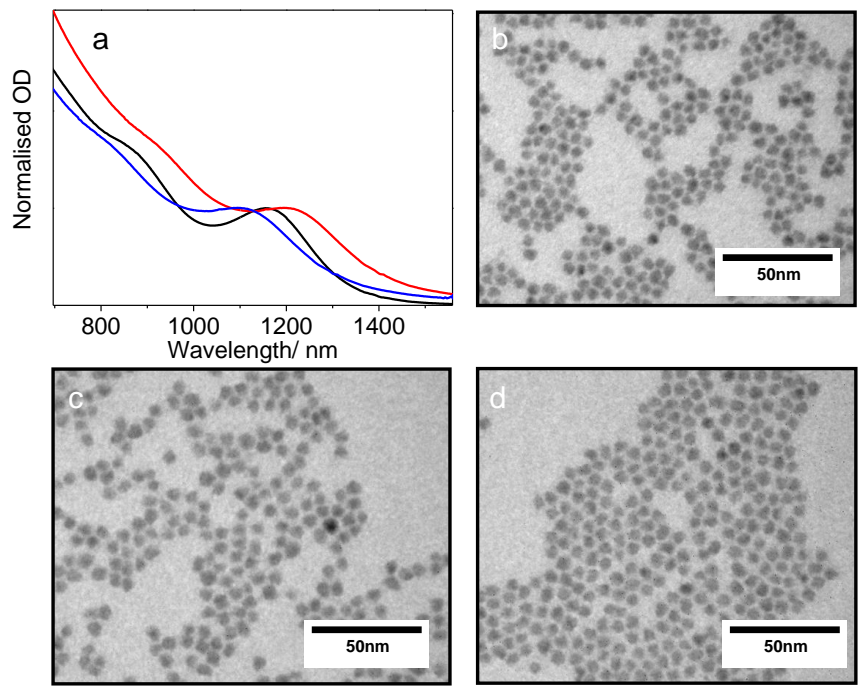

Figure S8. Absorption spectra and TEM images. (a) The absorption spectra of samples 1 (Control, Black), 2 (Cu:NC ratio of 400:1, Blue) and 3 (Ag:NC ratio of 190:1, Red). (b) TEM image of sample 1. (c) TEM image of sample 2. (d) TEM image of sample 3.

Structural characterization was carried out on 5nm diameter InAs NCs. These NCs were dissolved in toluene and divided into four samples. Sample 1 served as a control with a DDAB/DDA solution (with no metal salt) added to it. Cu solution was added to sample 2 resulting in a Cu:NC solution ratio of 3,900, which corresponds to 400 Cu atoms per NC after calibration. The addition of the Cu solution caused a 49nm (48meV) blue-shift in the first exciton feature. Ag solution was added to sample 3 resulting in a Ag:NC solution ratio of 1,100, which corresponds to 190 Ag atoms per NC after calibration. The addition of the Ag solution caused a 39nm (34meV) red-shift in the first exciton feature. Au solution was added to sample 4 resulting in a Au:NC solution ratio of 1,100, which corresponds to 130 Au atoms per NC after calibration. The addition of the Au solution resulted in a broadening of the first exciton peak leaving the position unchanged. The absorption spectra of samples 1-3 are shown in Figure S8a. These samples were precipitated and used for structural characterization. This included TEM imaging, sizing, and XRD and XPS measurements. As can be seen in Figure S8b-d metal domains on the surface of the NPs are not observed.

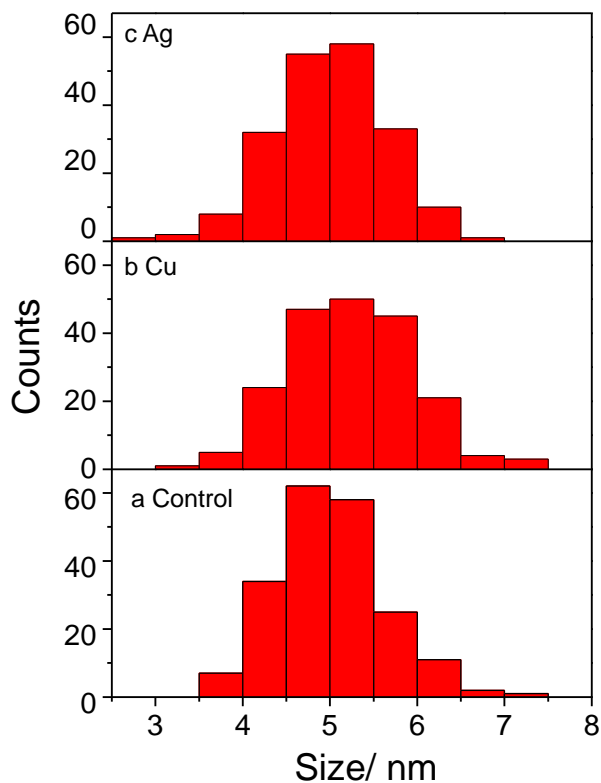

Figure S9. Size histograms. 200 TEM size measurements for samples 1-3. (a) Sample 1 the control sample Mean diameter 5.0nm, standard deviation (SD) 0.61nm. (b) Sample 2 (Cu:NC ratio of 400:1). Mean diameter 5.2nm, SD 0.69nm. (c) Sample 3 (Ag:NC ratio of 190:1). Mean diameter 5.0nm, SD 0.65nm.

These images were used to measure the size of the particles. 200 particles were measured for each sample. The size histograms are shown in Figure S9. Sample 1 (the control sample) has a mean diameter of 5.0nm and a standard deviation (SD) of 0.61nm, Figure S9a. Sample 2 (Cu:NC ratio of 400:1) has a mean diameter of 5.2nm and a SD of 0.69nm, Figure S9b. Sample 3 (Ag:NC ratio of 190:1) has a mean diameter of 5.0nm and a SD 0.65nm, Figure S9c.

## X-ray Powder Diffraction of Doped NCs

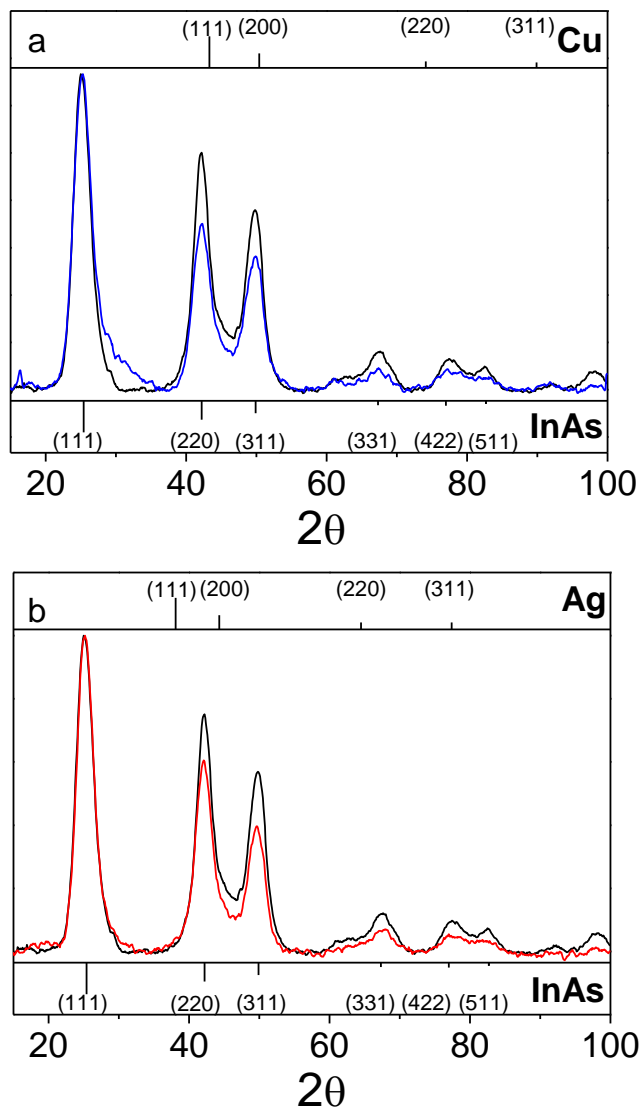

Figure S10. XRD diffractograms. (a) Samples 1 (control, Black) and sample 2 (Cu:NC 400:1, Blue). (b) Samples 1 (control, Black) and C (Ag:NC 190:1, Red).

Figure S10 shows the XRD diffractograms of samples 1-3. It is clearly seen that in both the Cu and Ag samples the main InAs reflections are present which indicates that the InAs crystal structure is generally preserved. It should be noted that evidence for disorder is seen in broadening of the peaks at higher angles, and also specifically for Cu, a tail appears around $2\theta = 30°$.

## The effect of the Impurity Atom on the Lattice Parameter

With regards to lattice distortions or changes as a result of substitutional doping, we analyze below the possible change in the lattice parameter. We shall apply the analysis to Ag, the largest impurity, as an upper limit of the changes. The addition of an impurity atom in a substitutional lattice site may result in the expansion or contraction of the lattice parameter, a. The change in the lattice parameter, $\Delta a$, is expressed by:

$$\frac{\Delta a}{a} = \beta N_i \qquad (1)$$

Where $N_i$ is the impurity concentration and $\beta$ is the expansion or contraction coefficient. $\beta$ can in turn be calculated using the covalent radii of the impurity, $r_i$, host atom, $r_h$, and the density of host atoms $N_h$ (4).

$$\beta = \left[ \frac{r_i - r_h}{r_h} \right] \frac{1}{N_h} \qquad (2)$$

Using the Pauli tetrahedral covalent radii along with the following values in conjunction with equations (1) and (2) at an assumed doping level of 10%:

$r_{Ag} = 1.52 \text{ Å} \quad r_{In} = 1.44 \text{ Å} \quad N_{InAs} = 1.8 \times 10^{22} \text{ atoms cm}^{-3} \quad N_{Ag} = 1.8 \times 10^{21} \text{ atoms cm}^{-3}$

$a_{InAs} = 6.06 \text{ Å}$

we obtain an expanded lattice constant of

$a_{In_{0.9}Ag_{0.1}As} = 6.10 \text{ Å}$

This is a difference of less than 1% from the original lattice parameter. The Debye-Scherrer broadening of the NC XRD signal will mask a shift in the reflection angle caused by such a small change in the lattice parameter, and we do not observe the shift in the XRD spectra.

## XPS Spectra of Doped NCs

Figure S11 shows the XPS survey spectra for samples 1-4. The In 3s, 3d, 3p and MNN as well as the As 3d, 3p and LMM peaks are clearly seen in all the spectra. In addition, the 1s and 2s peaks of O are observed. These arise from the $SiO_2$ of the Si wafer. The Cu 2p peaks are clearly visible in the spectra of sample 2 and absent from the control sample spectra.

The Ag 3p peaks are present in the spectra of sample 3 and absent from the control spectra. Similarly the Au 4f peaks are present in the spectra of sample 4 and absent from the control spectra.

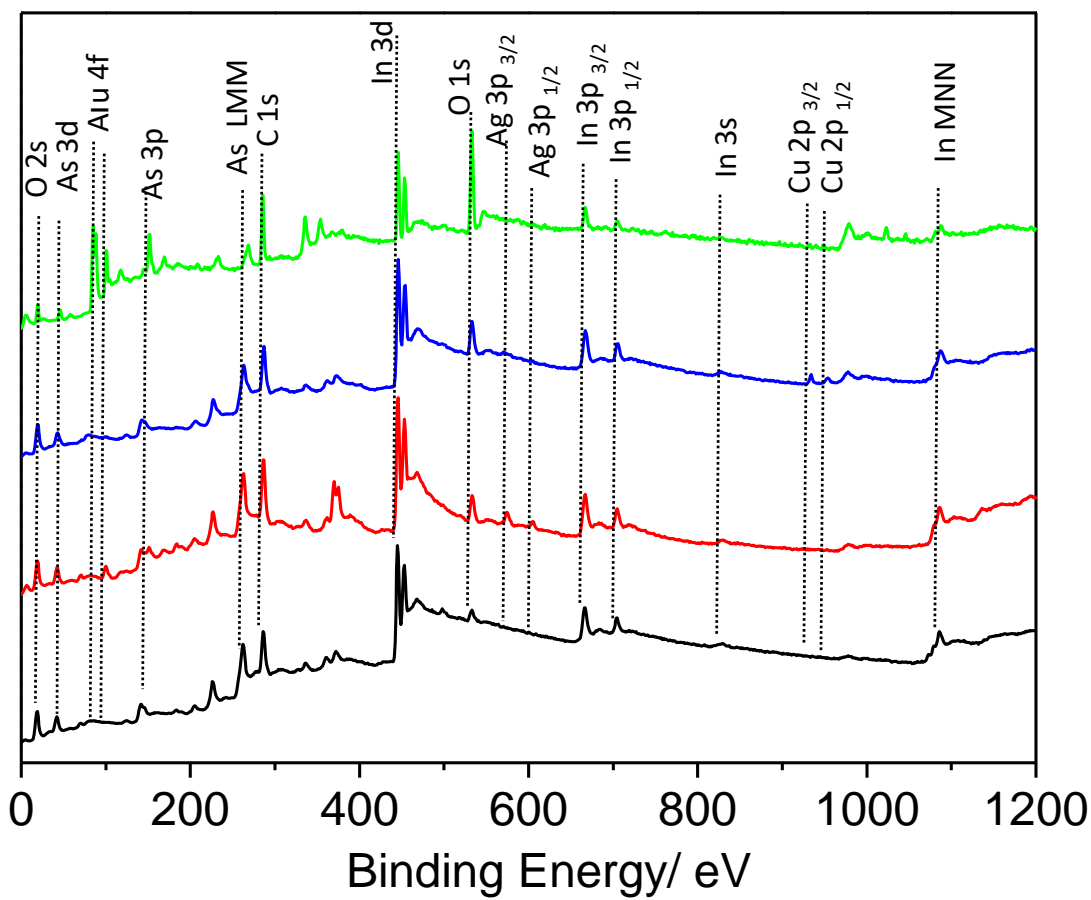

Figure S11. XPS Survey spectra. Sample 1 (control, Black, 5nm diameter), sample 2 (Cu:NC 400, Blue), sample 3 (Ag:NC 190, Red) and sample 4 (Au:NC 130, Green).

Figure S12 shows the high resolution XPS measurements of the various samples. The two peaks seen in Figure S12a correspond to the In $3d_{5/2}$ and $3d_{3/2}$ signals with binding energies of 444.2 and 451.7eV respectively. The peaks from the control, Ag and Au sample are almost identical whereas the peak for the Cu sample is broader. The peak fitting for the Cu doped sample (sample 2) shows contribution also from two additional slightly shifted peaks with binding energies of 444.8 and 452.3eV. These peaks may be related to the presence of a small amount of $In_2O_3$. The main peak in Figure S12b corresponds to the As 3d signal at 40.7eV. The control, Ag and Au samples are similar, whereas the Cu sample exhibits another smaller peak at a higher binding energy, 44.2eV. This secondary peak is assigned to $As_2O_3$. The presence of In and As oxides is noticeable as all the samples were exposed to air for the same time, less than five minutes, during loading into the vacuum chamber, yet only the Cu sample shows signs of oxidation. Figure S12c, focusing on the Ag peaks, shows for the Ag doped sample 3 the Ag $3d_{5/2}$ and $3d_{3/2}$ peaks with energies of 368.2 and 374.2eV, respectively. The peak positions match reports for both $Ag^0$ and $Ag^{1+}$, which are close in energy. In addition, there is a broad background signal at 371eV which is also observed in the control sample (this peak could not be identified and we assign it to a system artefact).

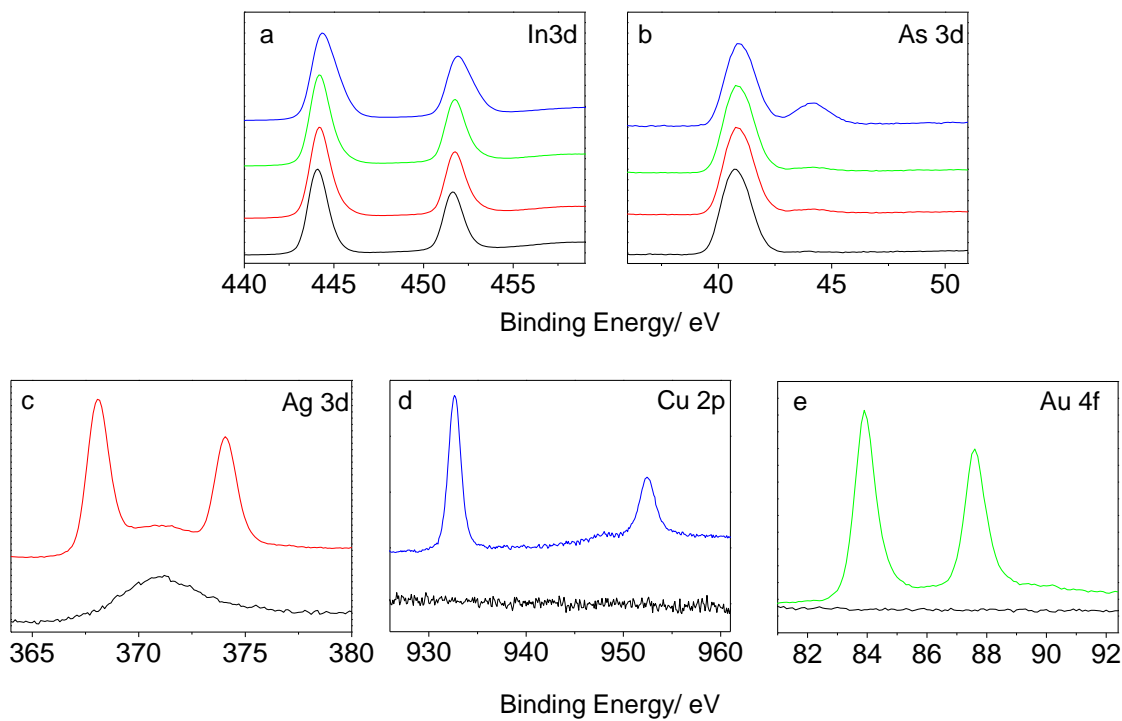

Figure S12. High resolution XPS spectra. In 3d (a), As 3d (b), Ag 3d (c) Cu 2p (d) and Au 4f (e) peaks for samples 1 (control, Black), sample 2 (Cu:NC 400, Blue), sample 3 (Ag:NC 190, Red) and sample 4 (Au:NC 130, Green).

The peaks exhibited in Figure S12d for the Cu sample are the Cu $2p_{3/2}$ and $2p_{1/2}$ with energies of 932.7 and 952.5eV respectively. The values match the reported binding energies for Cu in the 0 or +1 oxidation state, which are close in binding energy. The peaks exhibited in Figure S12e for the Au sample are the Au $4f_{7/2}$ and $4f_{5/2}$ peaks with energies of 83.9 and 87.5eV, matching reports for $Au^0$.

**Diffusion, Valence and Size Parameters of Cu, Ag & Au**

| Metal | Frequency Factor, $D_0$ (cm²/s) | Activation Energy, $E_a$ (eV) | Diffusion Constant, D, (cm²/s), @298K | $X_{24hrs}$ (nm) | 6-coordinate Ionic Radius (Å) | Oxidation States |
|-------|---------------------------------|-------------------------------|----------------------------------------|------------------|-------------------------------|------------------|
| Cu    | 3.6x10⁻³                        | 0.52                          | 5.8x10⁻¹²                              | 1.4x10⁴          | 0.73 (Cu⁺)                    | +1 or +2         |
| Ag    | 7.3x10⁻⁴                        | 0.26                          | 2.93x10⁻⁸                              | 1.0x10⁶          | 1.15 (Ag⁺)                    | +1               |
| Au    | 5.8x10⁻³                        | 0.65                          | 5.8x10⁻¹²                              | 1.4x10³          | 0.85 (Au³⁺)                   | +3               |

Table S1: Diffusion, valence and size parameters for Cu, Ag and Au (*5*).

The diffusion constants in Table S1 were extrapolated to room-temperature values using equation (1).

$$D = D_0 \exp(-E_a / RT) \tag{1}$$

The frequency factor, $D_0$, and the activation energy, $E_a$ were taken from literature values (**5**). *R* is the gas constant and, *T* is temperature. The extrapolated frequency factor was subsequently used to determine the root-mean-square distance that a diffusing atom would travel in 24 hours, $X_{24hrs}$, using equation (2).

$$X = \sqrt{4Dt} \tag{2}$$

Where *t* is time in seconds. As can be seen from Table S1, the $X_{24hrs}$ is much larger than the average size of the NC (2-5nm in diameter). Therefore a diffusing atom can penetrate into a NC in a very short time period. The six coordinate ionic radii are reported from literature values (**5**) as they are readily available for all the relevant species. For comparison, the value for $In^{3+}$ is 0.8 Å.

**Scanning Tunneling Microscopy and Spectroscopy**

For the STM measurements, the nanocrystals were deposited on a flat Au(111) film by letting a drop of toluene-nanocrystal solution slowly dry on the substrate. The samples were mounted inside a homemade cryogenic STM that was promptly evacuated and cooled down to 4.2 K for data acquisition, so that samples were exposed to ambient air for less than 15 min. Spatially-resolved tunneling spectra (dI/dV vs. V characteristics, proportional to the local density of states), were measured by first positioning the tip above a single nanocrystal, thus forming a double barrier tunnel junction. Care was taken to retract the STM tip as much as possible from the NC in order to minimize the effect of voltage division between the junctions, which tends to broaden the measured energy-gap and level spacings. However, it is important to note that the main trends reported for the shifts in the center of the gap did not depend on the STM (current and voltage) settings. Four different Au-doped InAs NCs, with a nominal doping level of 77 atoms per NC were measured (from the same sample). Eight different Cu-doped InAs NCs, with a nominal level of doping of 160 Cu atoms per NC were measured. Ten different Ag-doped NCs, with two nominal levels of doping, of 15 and 60 Ag atoms per NC, were measured.

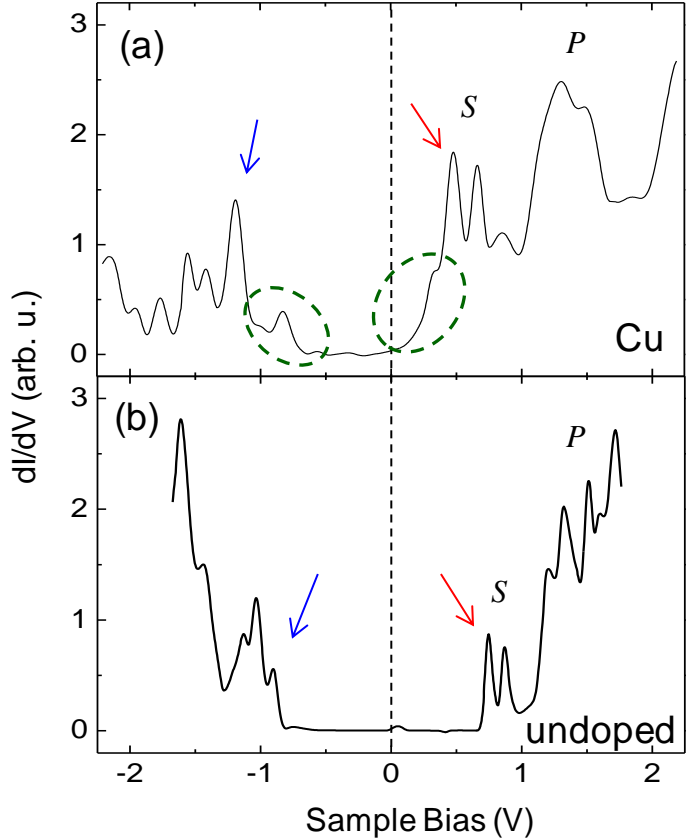

Figure S13. Tunneling spectra. At 4.2K for Cu-doped (a) and undoped (b) InAs quantum-dots, 4.2 nm in diameter. The blue and red arrows mark the onsets of the valence and conduction bands, respectively, while the dashed green ellipses mark tail and localized in-gap states. At positive bias the 1Se doublet (S) and the 1Pe multiplet can be observed, although the latter is not well resolved for the doped nanocrystal.

In Figure S13 we present the tunneling spectra for undoped and Cu-doped InAs quantum-dots, presented in Figure 3 of the paper, but over the full voltage range. The onsets of the conduction and valence bands as well as the tail and localized in-gap states are marked.

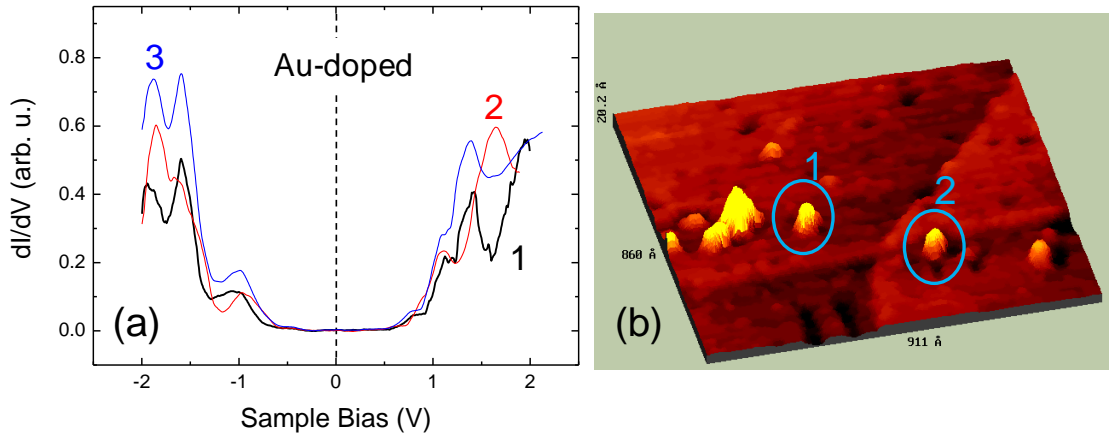

Figure S14: (a) Tunneling spectra measured on three Au doped InAs NCs with a nominal doping level of 77 Au atoms per NC. (b) Topographic images for two such NCs.

Additional tunneling spectra are presented here for each one of the systems studied. These cover the whole voltage range and not only the bandgap region, as shown in the paper. Spectra and topographic images for Au doped InAs NCs with a nominal doping level of 77 Au atoms per NC are presented in Figure S14. The apparent energy gaps vary slightly from one spectrum to another and the (broadened) excited level structure fluctuate to even a larger extent, but nevertheless the data clearly demonstrate that no shift with respect to zero bias takes place for any of these spectra.

Spectra for Cu doped InAs NCs with a nominal level of doping of 160 Cu atoms per NC are shown in Figure S15. All the spectra clearly show a similar shift of the bandgap towards negative bias. Spectra for Ag doped InAs NCs with nominal doping levels of 15 and 60 Ag atoms per QD are presented in Figure S16. All clearly exhibit a shift of the bandgap towards positive bias. The effect appears to be slightly larger in the case of higher doping, but obviously to a marginal extent considering the fluctuations in the measured level structure from one NC to another. This data further demonstrates and establishes the conclusions drawn from figure 3 of the paper, that the bandgap, or conduction and valance band edges, shift to positive energies with Ag doping, to negative energies upon Cu doping, and remain nearly intact (with respect to the undoped InAs NCs) for Au doping. This effect is far larger than the variations observed in the spectra from one NC to another. The effect of the degree of doping, however, could not be clearly discerned due to the broadening of the level structure upon doping, and also the uncertainty in the actual doping level for each one of the single NCs studied, which obviously fluctuates with respect to the average-nominal doping level of the sample.

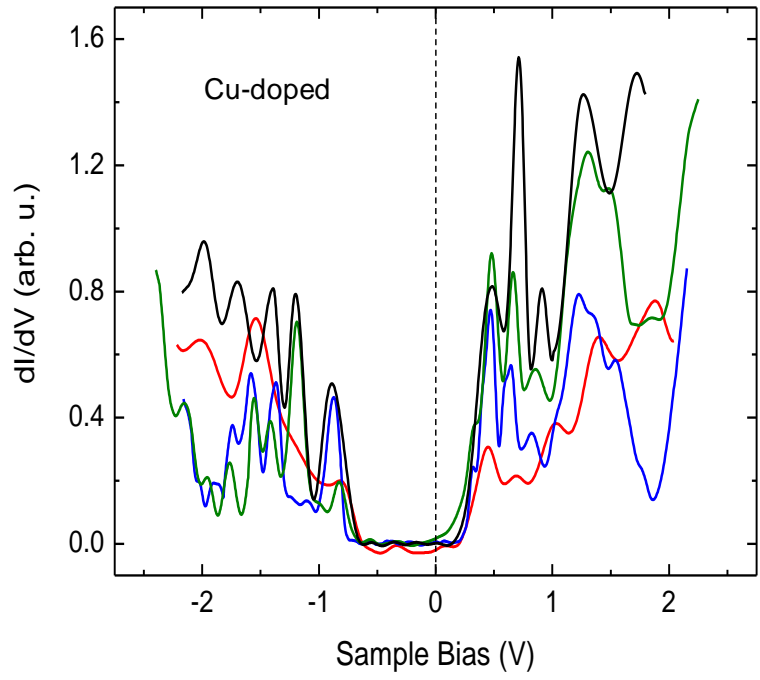

Figure S15: Tunneling spectra measured on four Cu doped InAs NCs with a nominal doping level of 160 Cu atoms per NC.

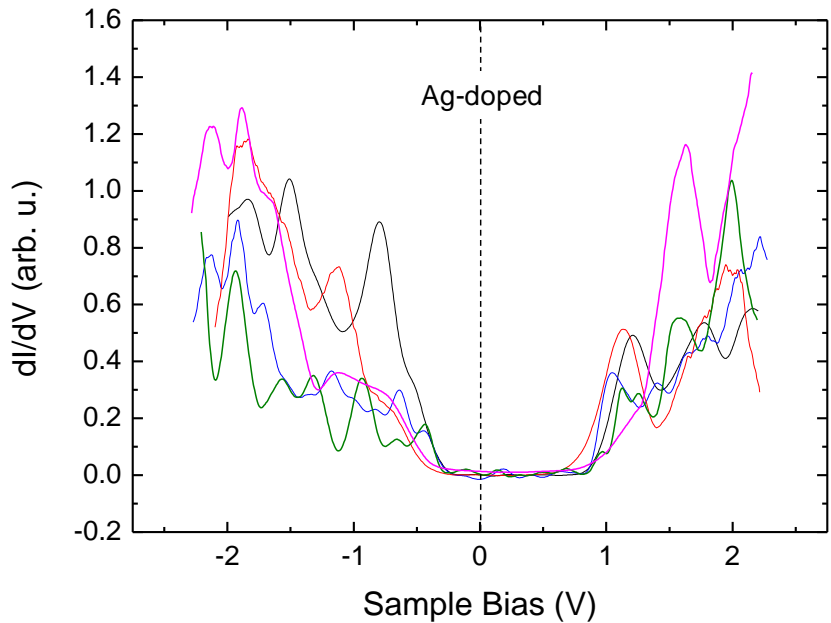

Figure S16: Tunneling spectra measured on Ag doped InAs NCs, two from a sample of lower nominal doping, 15 Ag atoms per NC, (red and magenta curves) and three from a sample of higher doping, nominally 60 Ag atoms per NC, (black, green and blue).

**Chemical and Structural Considerations of Doped InAs Nanocrystals**

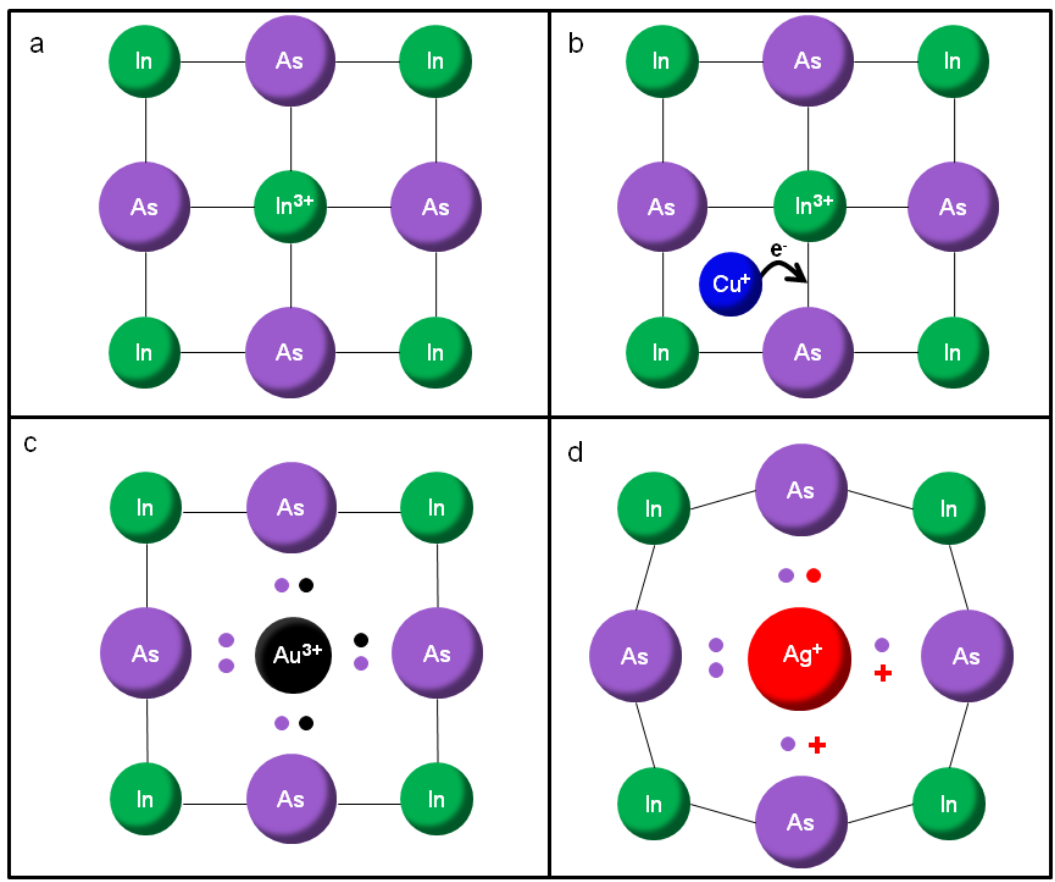

Figure S17. Lewis structure diagrams which depict a simplified view of the bonding in the InAs lattice. Bonding electrons are represented either by black lines or by paired dots, whose color matches that of the atom to which it belongs. Plus signs indicate the lack of an electron in a bonding orbital. (a) A 2D simplified representation of the InAs lattice with tetrahedral bonding where the central In atom has a formal +3 oxidation state and eight bonding electrons. (b) InAs lattice with a Cu atom at an interstitial site donating valence electrons to the crystal leading to *n*-type doping. The lattice distortion expected in this case has been omitted here due to its small size. (c) A substitutional Au impurity on an In site in the InAs lattice. Lattice distortion is small due to the similar size of In and Au. The three Au valence electrons are able to participate in bonding to adjacent atoms, as occurs with In atoms, thus leaving the sample electronically undoped. (d) A substitutional Ag impurity on an In site in the InAs lattice. The Ag causes lattice distortion due to its larger size and introduces two acceptor sites into the lattice due to its lack of valence electrons leading to *p*-type doping. Note that the bonding in InAs structure has covalent-ionic character and this is not fully described here. The aim of the schemes is to provide a chemically intuitive framework for the doping effects.

We consider, within a simplified Lewis structure framework, the effects of impurity size, location and the number of valence electrons on the interaction between Cu, Ag and Au and the InAs lattice (Figure S17). Cu is small (Table S1) and takes up an interstitial site leading to a small lattice distortion. Such an interstitial Cu will tend to partly donate its valence electrons to the QD leading to *n*-type doping as depicted in Figure S17b. Indeed Cu is known to be an interstitial donor in InAs (*6-9*). Au is known to be a substitutional impurity in Si and III-V semiconductors. It has an isovalent oxidation state with In, Table S1. It therefore acts as an isovalent impurity which does not introduce charge carriers to the lattice. In addition it has a similar size to In (Table S1) and so does not distort the lattice significantly, as shown in Figure S17c. Ag, like Au, is known to be a substitutional impurity in Si and III-V semiconductors. Its tendency for a +1 oxidation state makes it an acceptor leading to *p*-type doping, and its large size causes significant lattice distortion as shown in Figure S17d.

The size of an impurity can influence its positioning within the lattice. Impurities with small sizes are more likely to take up interstitial sites whereas larger impurities tend to accommodate substitutional sites. The location of the impurity within the lattice affects the type of interaction that occurs. The location of the impurity, in conjunction with its size, affect the extent of the lattice distortion. Calculations on the distortion in bulk silicon caused by a substitutional Cu impurity show that the change in bond length of the first shell of neighboring atoms is around 10% (*10*). Similar calculations for interstitial first-row transition metals indicate that the change in bond length is 3.6% (*11*).

This analysis leads to a conclusion that distortions caused by the Ag substitutional impurity are much larger than those caused by the Cu interstitial atom. Although the above mentioned calculations on the distortion of bond lengths for interstitial transition metals in silicon do not treat Cu, it is smaller than the metals considered, and hence it is expected to lead to a smaller change in bond length of less than 3.6%. In addition, InAs has a larger lattice constant than silicon and this should result in a further reduction of the distortion. On the other hand Ag is significantly larger than Cu and therefore it is expected that a substitutional Ag impurity would result in a change in bond length greater than 10%.

# Theory of Heavily Doped Nanocrystals

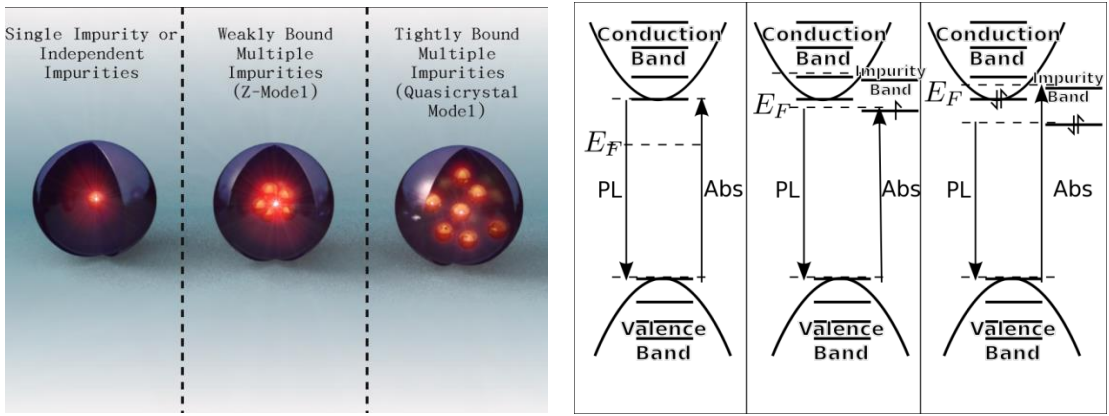

Figure S18. (left) An illustration of the three models investigated. (right) An illustration of how (for example, *n*-type) doping of a nanocrystal can result in both red and blue shifts in optical spectrum measurements: the intrinsic crystal is shown on the left. To the center, a single donor impurity adds a free level below the intrinsic Fermi energy, causing red shifts in both PL and absorption spectrum measurements. To the right, several impurities packed close together within the crystal may have lower energy levels than the single or isolated impurity, but the electronic ground state may be such that the original Fermi level is occupied as a result of the doping. This leads to a blue shift in the absorption spectrum along with a red shift in the PL.

In this section we provide a detailed discussion of the theories developed to analyze and understand the experimental results. This is done through the consideration of three separate models of nanocrystal doping corresponding to three distinct physical limits, only one of which is directly relevant to the experiments at hand. This complete picture is useful in making a convincing argument that doping is indeed observed. It is worth noting that each of these models correctly scales to the bulk case, where confinement is not an issue and the physics is well known.

In the literature, highly concentrated impurities in bulk have been treated extensively (*12*), and more recently some properties of hydrogenic dopants within quantum dots of various geometries have also been addressed by both approximate and exact methods (*13-18*). However, no theoretical framework to date has allowed for even a conceptual discussion of the energetic and optical properties of excitations in the interesting regime where multiple impurities are confined within a nanocrystal. To remedy this oversight, several simple limits of such a system are considered in the text below. The left panel in Figure S18 presents the models - each of which will be discussed in detail later in the text. The right panel of the same figure provides an explanation of how doping might modify the physical picture.

## Single Hydrogenic Impurity

The simplest case, and the only one to have been studied extensively, is that of a single impurity somewhere within the crystal. One phrasing of this problem which can be reduced analytically to that of a 1D root search is a hydrogenic impurity at the center of a spherical finite well (*18*). The hydrogenic approximation is crude at best, as full electrostatic screening does not form on the length scales in question - nevertheless, it is a good starting point. In essence, one must solve the spherically symmetric Schrödinger equation:

$$\left( \frac{\mathbf{p}^2}{2m^*(\mathbf{r})} - \frac{e^2 z}{\mathtt{y}(\mathbf{r})|\mathbf{r}|} + V(\mathbf{r}) \right) \psi(\mathbf{r}) = E\psi(\mathbf{r})$$

$$V(\mathbf{r}) = \begin{cases} 0 & |\mathbf{r}| \leq R, \\ U & |\mathbf{r}| > R. \end{cases}$$

(3)

Note that the dielectric constant $\mathtt{y}(\mathbf{r})$ and effective mass $m^*(\mathbf{r})$ also vary in space. Specifically, they are different inside and outside the crystal:

$$\mathtt{y}(\mathbf{r}) = \begin{cases} \mathtt{y} & |\mathbf{r}| \leq R \\ \mathtt{y}' & |\mathbf{r}| > R \end{cases}$$

$$m^*(\mathbf{r}) = \begin{cases} m & |\mathbf{r}| \leq R \\ m' & |\mathbf{r}| > R \end{cases}$$

(4)

Within each region, the solutions are described in spherical coordinates by $\psi_{n\ell m}(\mathbf{r}) = Y_\ell^m(\vartheta,\varphi) R_{n\ell}(r)$. By neglecting the Coulomb interaction one obtains the "bare" problem, where due to the boundary conditions at $r \to 0$ and $r \to \infty$ the radial dependence of the wavefunction has the form

$$R_{n\ell}^b = \begin{cases} a_{n\ell} j_\ell(k_{n\ell} r) & r \leq R \\ a'_{n\ell} f_\ell(\kappa_{n\ell} r) & r > R \end{cases}$$

(5)

Here, $j_\ell(x) = \sqrt{\frac{\pi}{2x}} J_{\ell+1/2}(x)$ is the $\ell^{\text{th}}$ order spherical Bessel function of the first kind and $f_\ell(x) = \sqrt{\frac{\pi}{2x}} K_{\ell+1/2}(x)$ is the analogously defined spherical modified Bessel function of the second kind. Similarly, when the impurity is taken into account the radial dependence is

$$R^b_{n\ell}(r) = \begin{cases} a'_{n\ell} w^i_\ell(k_{n\ell}r) \equiv a_{n\ell} e^{-\frac{k_{n\ell}r}{2}} (k_{n\ell}r)^\ell F(\ell+1-\lambda, 2\ell+2, kr) & r \leq R \\ a'_{n\ell} w^o_\ell(\kappa_{n\ell}r) \equiv a'_{n\ell} e^{-\frac{\kappa_{n\ell}r}{2}} (\kappa_{n\ell}r)^\ell U(\ell+1-\lambda', 2\ell+2, \kappa r) & r > R \end{cases} \quad (6)$$

With $\lambda = \frac{2mze^2}{\hbar^2 \varepsilon k}$ and $\lambda' = \frac{2m'ze^2}{\hbar^2 \varepsilon' \kappa}$. $F$ and $U$ are the confluent hypergeometric functions of the first and second kind, respectively. $k$ and $\kappa$ are related through their relation to the total energy, since

$$E = \frac{\hbar^2 k^2}{2m} = \frac{\hbar^2 \kappa^2}{2m'} + U. \qquad (7)$$

Due to the spatial dependence of the effective mass, an erroneous non-Hermitian term has actually been introduced into the Hamiltonian: a more rigorous treatment requires the kinetic energy term be replaced by the Hermitian Ben-Daniel-Duke Hamiltonian $\mathbf{p} \frac{1}{2m^*(\mathbf{r})} \mathbf{p}$, which in this case amounts to taking boundary conditions such that the probability current rather than the derivative of the wavefunction is conserved across the interface[19]. Demanding also that the wavefunction be continuous then allows the boundary conditions to be written as:

$$\begin{cases} R_{n\ell}(kr)\big|_{r \to R^-} = R_{n\ell}(\kappa r)\big|_{r \to R^+} \\ \frac{1}{m} \partial_r R_{n\ell}(kr)\big|_{r \to R^-} = \frac{1}{m'} \partial_r R_{n\ell}(\kappa r)\big|_{r \to R^+} \end{cases} \qquad (8)$$

or

$$\frac{\partial_r R_{n\ell}(kr)}{R_{n\ell}(kr)}\bigg|_{r \to R^-} - \frac{m}{m'} \frac{\partial_r R_{n\ell}(\kappa r)}{R_{n\ell}(\kappa r)}\bigg|_{r \to R^+} = 0 \qquad (9)$$

The roots of this equation determine the bound state energies and allow the explicit construction of the exact wavefunctions. This model is in principle appropriate for describing both electrons in the conduction band and holes in the valence band, as well as both donor and acceptor impurities, as long as the relevant parameters are utilized.

One is often interested in the binding energy, which may be defined as the difference between the energy of a state $i$ in the presence of an impurity and the energy of the corresponding state in the undoped dot,

$$E_b^i = E_{\text{imp}}^i - E_{\text{dot}}^i. \tag{10}$$

Specifically, the shift $\Delta$ in the absorption spectrum edge is equal to $E_b^0$. Variational arguments may be used to make it clear that for the ground state, some binding occurs with the addition of any attractive potential (the binding energy is always negative for donors and positive for acceptors). As a result, this model can only account for red shifts in both spectra, when considering both donor and acceptor impurities. It remains useful for determining to first order the effect of parameters such as the crystal size, the barrier energy, the band effective mass and the effective dielectric ratio on the level structure of the crystal.

Note that we have completely neglected the interaction energy between the electron-hole pair and the impurity. To lowest order perturbation treatment in the Coulomb coupling, this interaction is given by $E_c = E_{eh} + E_{ih} + E_{ie}$, where $E_{eh}$ is the electron-hole interaction (direct and exchange), $E_{ih}$ is the impurity-hole interaction, and $E_{ie}$ is the impurity-electron interaction. $E_c$ should be added to $E_{\text{imp}}^i$. Similarly, for the exciton energy of the dot, $E_{\text{dot}}^i$, one should add the electron-hole interaction, $E_{eh}$ (the only interaction in the undoped QD) and thus, $E_b^i = E_{\text{imp}}^i - E_{\text{dot}}^i \rightarrow E_{\text{imp}}^i - E_{\text{dot}}^i + E_{ih} + E_{ie}$. These additional terms are rather small, but there is another factor which when taken into account reduces its importance even more greatly: the interaction between the pair electron and the impurity electron is almost exactly cancelled out by the interaction between the pair hole and the impurity electron, i.e., $E_{ih} \approx -E_{ie}$. Nevertheless, it is simple to numerically evaluate the interaction within perturbation theory, and we have found that it makes essentially no difference.

With the exact wave functions known, it is simple enough to calculate the dipole absorption spectrum of the nanocrystal (*20*). This is done for a certain set of parameters in Figure S19.

The discussion above is suitable not only to the description of a crystal with a single impurity, but also to the case where multiple impurity carriers are entirely independent of each other and of other impurity ions. For $N$ impurities, $N$ perfectly degenerate half-filled impurity ground states occur. This is not a particularly physical situation, especially in a small dot in which the impurity is assumed to be exactly at the center; however, it is illuminating to consider the fact that as the perfect independence is broken, interaction between impurities will begin to remove the degeneracies between levels. If enough of the levels are raised above the intrinsic ground state by this, such that it becomes filled, blue shifts in the absorption spectrum may occur. The next section takes this idea to a more quantitative level.

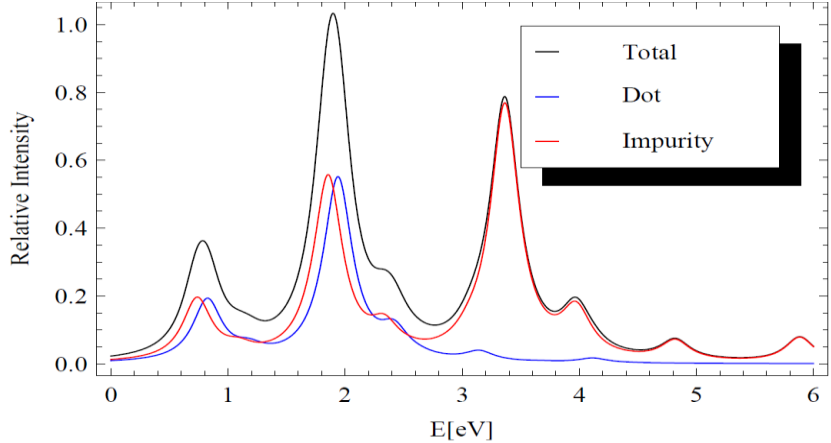

Figure S19. Theoretical spectrum. The effect of doping on the absorption spectrum in the Golden Rule dipole approximation is shown as calculated for a single donor impurity in the center of an InAs nanocrystal with $R = 2.5\text{nm}$ and assuming $U = 3\text{eV}$. The relative spectral intensity as a function of the excitation energy of the doped particle (in red) is compared to that of the undoped particle (in black). A red shift of $\sim 80\text{meV}$ is clearly visible, as well as the appearance of further peaks at higher energies.

## Multiple Strongly Interacting Impurities: The Z-model

The impurity wave functions found in the previous subsection are strongly delocalized within the nanocrystal. If the impurity ions can be assumed to be small and located near the center of the crystal, such that the donor electrons or acceptor holes see their combined and screened charge, it seems reasonable to model the impurities by a $z > 1$ Coulomb term. Of course, this picture also holds when considering multivalent impurities. One must, as a first approximation, take into account that the single-electron levels are filled in accordance with the Pauli principle, or for greater accuracy consider some form of the full many-body problem. For the purpose of this argument the first approach is taken, and it is later argued that neglecting many-body effects is justified as long as one is satisfied with a qualitative description in the strong confinement regime, where the energetics are dominated by the kinetic term.

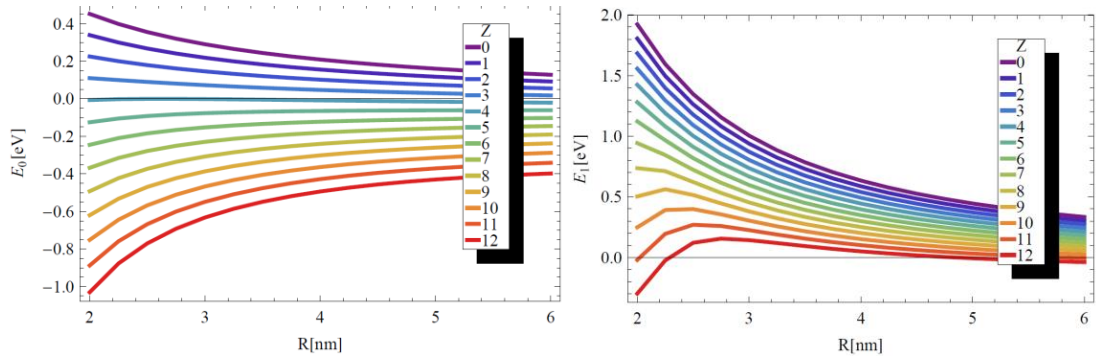

Figure S20. Z-model energy levels. (left) First and (right) second energy levels in the Z-model, for different values of Z as a function of the nanocrystal radius. The parameters are chosen for InAs. Specifically, the effective electron mass within the particle is 0.027 $m_e$, the inner dielectric constant $\epsilon$ is 12, the outer dielectric constant $\epsilon'$ is 2 and, as before, the barrier height is taken to be 3eV. Zero energy corresponds to the bottom of the conduction band. Note that since the shift is defined in relation to the $Z=0$ case, in this case the first level's shift is zero by definition.

The first and second energy levels as calculated within the Z-model are shown in Figure S20. Clearly and as expected, these levels are reduced by increasing the ionic valency and therefore the depth of the central Coulomb well. It is also apparent that when confinement is increased, this effect grows appreciably stronger. One can therefore immediately expect larger and more varied shifts when the NCs are decreased in size.

When the level degeneracy is correctly taken into account, one can deduce the shift in the absorption spectrum from the shift in the Fermi level as compared to the dot ground state. The result for InAs is displayed in Figure S21 (left): for $Z=1$ this is identical to the single-impurity case. For $Z=2-3$ no shifts occur since the Fermi level is at the dot ground state, as shown in Figure S21 (right). At still higher values, blue shifts of various magnitudes occur. All shifts become substantially smaller in the absence of confinement.

Perhaps the most striking feature of the Z-model is the immediate appearance of a blue shift at very low impurity concentrations, and its decrease in magnitude when more impurities are added. This is a result of the unbroken spherical symmetry of the model, which leads to high level degeneracy. While this limit is certainly of theoretical interest, it clearly does not reflect well the experimental results presented here.

Note that the effect of interactions between electrons has not been taken into account in this context. When one considers a large number of electrons in a high-valence impurity, one cannot, of course, continue to justify this: a quantitative treatment certainly requires a more sophisticated many-body approach. Qualitatively, one can say that an extra positive term $\sim Z^2$ will be added to the impurity. This is expected to significantly modify the system's behaviour at large $Z$ and small $R$ in such a way that the blue shift effect we have discussed will increase at these parameters; the interesting decrease with $Z$ we have predicted may or may not disappear, depending on the detailed magnitude of the effect. From our perturbative estimation of the interaction energy, we expect some of the effect will survive, but further exploration will be needed to provide a definite answer.

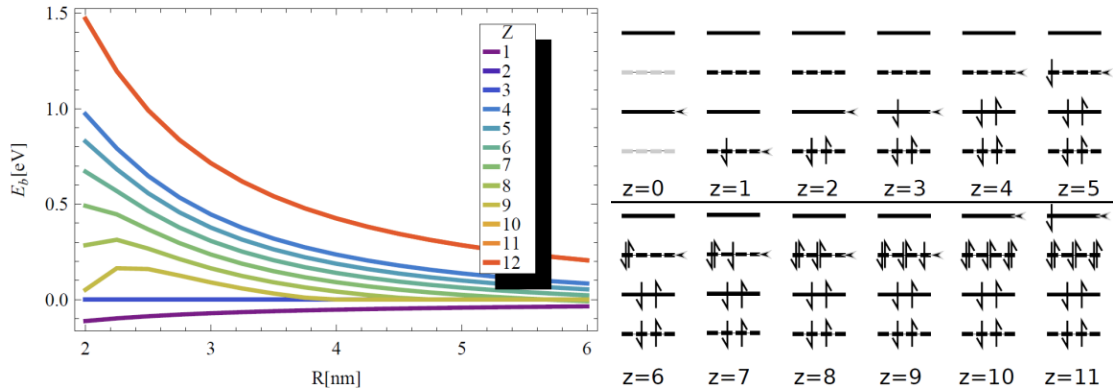

Figure S21. Z-model energy shifts. (left) Shifts in the absorption spectrum edge $\Delta$ as a function of nanocrystal radius at several values of the valency $Z$. (right) For comparison, an illustration of the level occupation for $Z=1-11$ assuming that the system is described by two s-levels (dotted line impurity lowest, then full line dot) followed by two p-levels (in the same order). The small arrows to the left of the level lines, point to the lowest unoccupied orbital. See the discussion in the text for more details.

## Multiple Scattered Impurities: the Pseudo-crystal Model

While the previous limits discussed provide interesting insight into the physics of electrons within doped nanocrystals, they both tacitly assume that screening occurs on a scale so small that the impurities may be described as point charges within a dielectric medium. In the more familiar context of impurities within bulk crystals, including the heavily doped regime, this makes sense: even if the impurities are close on the scale of their effective Bohr radii, they remain extremely far away on the atomic scale. However, in semiconductor nanocrystals this is not the case, and the hydrogenic approximation fails (*21*). While it is safe to assume that the energy of impurity levels remains largely independent on the impurity location (*21,22*), assuming that the wave function has the form of equation (6) rather than some more complicated form dependent on the specific chemistry of the system is most likely wrong. Therefore, either an atomistic treatment or a phenomenological one with this in mind is called for.

This section describes a phenomenological model aimed at this problem: we parameterized the properties of randomly distributed impurity levels and combined them into a pseudo-band by way of the tight-binding method, with hydrogen-like base states of energy $\varepsilon$:

$$H_{ij} = \delta_{ij}\varepsilon + (1-\delta_{ij})\gamma_{ij}$$
$$\gamma_{ij} = -2\gamma e^{-r_{ij}/a}$$
(11)

The parameter $\varepsilon$ is the impurity binding energy (which at large particle radii goes to the bulk value). Here, we take $\varepsilon$ to be constant independent of the position of the impurity within the QD (*23*). The parameters $\gamma$ and $a$ describe, respectively, the overlap energy at short distances and the typical length scale at which state overlap disappears. In bulk $\gamma = \varepsilon$ and $a = \dfrac{\hbar^2 \mathfrak{y}}{m^* e^2}$ is determined by the screening and the effective mass ($m^*$ is the effective mass of the carrier and $\mathfrak{y}$ is the dielectric constant), but we will allow these parameters to vary more freely. This variation will be considered a heuristic way of incorporating the effects of confinement into the model. By diagonalizing the Hamiltonian above and populating the impurity band and dot levels, we obtain a shifted LUMO level energy. This is exactly the shift $\Delta$ in the absorption spectrum (since the zero of the energy scale corresponds to the dot binding energy).

The above model is, therefore, that of "deep impurities". The impurity electrons (or holes) are bound strongly to the impurity ions, yet are delocalized enough that mixed states can form. This formalism is actually similar to the model used in the heavily doped bulk regime (*12*), but the justification and parameters are entirely different except in that regime (to which it returns in the appropriate limits). On the other hand, in the infinite dilution limit (a single impurity) and with perfect screening restored, one obviously returns to the case of hydrogenic impurities described in the previous subsections. It is the intermediate regime which is of greatest interest here, as it most closely corresponds to the experiments at hand.

While it is perhaps simplest and cleanest to perform the calculation described above for samples which have a well-defined integer number of impurities within them, this is not a realistic assumption in the experimental scenario discussed here, where some distribution of impurity numbers must exist within the measured ensemble. One could assume a Gaussian distribution of the impurity number around the average, for instance - but since the variance is unknown and expected to be large compared to the average, we have preferred to use a Poisson distribution, which has the advantage of requiring no further parameters. The specific choice does not affect the results qualitatively, as long as the distribution is wide enough to smooth over the discontinuous nature of the shift at small impurity concentrations. Poisson distribution is also consistent with a diffusive mechanism of impurities into the nanocrystals.

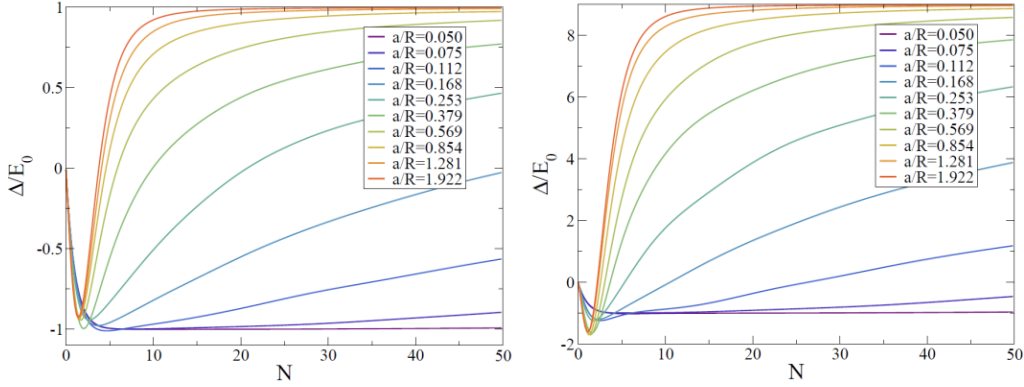

Figure S22. The average absorption shift $\Delta$ in the pseudocrystal model is plotted as a function of the average number of randomly distributed impurities within the crystal. The actual number of impurities for each crystal was averaged over a Poisson distribution (see the text). The InAs levels were calculated from an effective mass model with a finite barrier with a finite barrier $U = 3\text{eV}$, electron effective mass $m_e^* = 0.027 m_e$, hole effective mass $m_h^* = 0.41 m_e$, and dielectric constant $\varepsilon = 12$. In the left plot, the short-range mixing energy $\gamma = \varepsilon$ as in the bulk limit, while in the right plot $\gamma = 5\varepsilon$.

In practice, the results of the model as plotted in Figure S22 closely resemble those observed when Cu impurities are introduced if one takes $\gamma \approx E_0$ and $a \approx R$. While fits may be made, the quality of the data is such that they would be of little physical significance. It can, however, be stated with some confidence that if the model is taken to be correct and is used to reproduce the experiments, impurity electrons are highly localized (the impurities are indeed "deep") and mix strongly when impurities are close to each other.

## Band Tailing in Nanocrystals

A concern one might raise when optical measurements are made to detect doping is that any modification in the spectral properties of the nanoparticles is due to distortion of the symmetric crystal structure by the dopants, rather than by the donation or acceptance of electrons and holes. In the bulk, this is termed as "band tailing" and is known to account for red shifts in the spectrum since it leads to broadening and shifting of the density of states into the gap region. However, the implication of disorder for nanocrystals is still an open question.

In Figure S23 we show the calculated density of states (DOS) for a series of InAs nanocrystals at different levels of disorder. The electronic structure of the nanocrystal was described within the real-space pseudopotential method (24,25). The local screened pseudopotentials used in the results shown here were fitted to reproduce the experimental bulk band-gap and effective masses for InAs (26) neglecting spin orbit coupling (27). Ligand potentials were used to represent the passivation layer (24). We employed a Monte Carlo method (28) which computes directly the DOS as $p^{-1} \text{Im} Tr \left[ (E - H + ih)^{-1} \right]$ by expanding the Green's function in a Newton interpolation polynomial[29] (for the results shown, $h = 0.05 eV$). The trace is then preformed as a sum over random wave functions $y_i$ with values of $\pm 1$: $p^{-1} \text{Im} Tr \left[ (E - H + ih)^{-1} \right] = \frac{1}{pN} \text{Im} \sum_{i=1}^{N} \langle y_i | (E - H + ih)^{-1} | y_i \rangle$. The size of the grid to represent $y_i$ depends on the size of the nanocrystals. Grids were $32 \times 32 \times 32$, $48 \times 48 \times 48$, $64 \times 64 \times 64$, $72 \times 72 \times 72$, and $100 \times 100 \times 100$ for $In_{19}As_{16}$, $In_{44}As_{43}$, $In_{140}As_{141}$, $In_{264}As_{249}$, and $In_{652}As_{627}$, respectively. To reduce the computational time, Monte Carlo averaging over different realizations of disorder were carried parallel to the aforementioned Monte Carlo trace.

To describe the effect of band tailing induced by doping we randomly chose an In/As atom (or both) and displaced it in a random direction. A normal distribution was used for the magnitude of displacements with a width taken as a parameter. We repeated this procedure for a number of In/As atoms (or both) in the nanocrystal. The level of disorder was then determined by the number of displaced atoms relative to the total number of atoms in the nanocrystal (not including the passivation layer). For the results shown in Figure S23 we used a width of 10% the In-As bond length in the bulk. Similar results with smaller shifts in the DOS were obtained for smaller widths. A collapse of the band-structure was obtained for width of ~20% the In-As bond length.

From the DOS we can calculate the average shifts. Since the conduction band shows very small shifts as a result of the light effective mass of the electron, we focus on the shifts observed for the hole near the top of the valance band. The average shifts shown in Figure S24 were calculated by averaging the local shift between the DOS of the ordered nanocrystal and the DOS of the disordered one and averaging the local shift over the range of energies starting at the isosbestic point to energies slightly below the Fermi level, where the DOS of the ordered nanocrystals has practically decayed to 0. The shifts calculated this way for the displacement of the In atoms (upper panel of Figure S24) are slightly smaller than the shifts calculated for the displacement of As atoms (lower panel of Figure S24). More importantly, the shifts increase with decreasing nanocrystal size (in agreement with the experimental observation) and saturate at lower disorder levels for larger nanocrystals (also consistent with experimental observations).

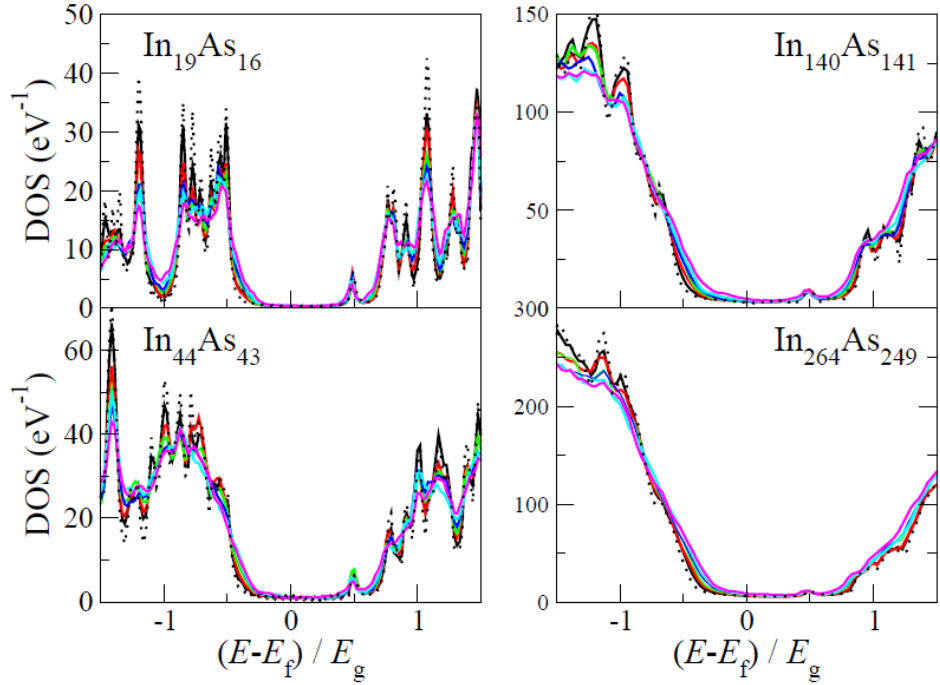

Figure S23. The effect of disorder on the DOS. The density of states (DOS) for different nanocrystals as a function of energy for different levels of disorder. Solid black, red, green, blue, cyan, and magenta are for disorder levels of 2.8%, 5.7%, 8.6%, 11.4%, 14.3% and 17.1%, respectively. Dashed black curve is the density of states with no disorder. The energy is scaled with respect to the band gap of the ordered nanocrystal for each size respectively. The Fermi energy, $E_f$, is taken mid way between the HOMO and LUMO and $E_g$ is the undoped quasi-particle gap. This sets the top of the valance band to be at -½ and the bottom the conduction band at +½.

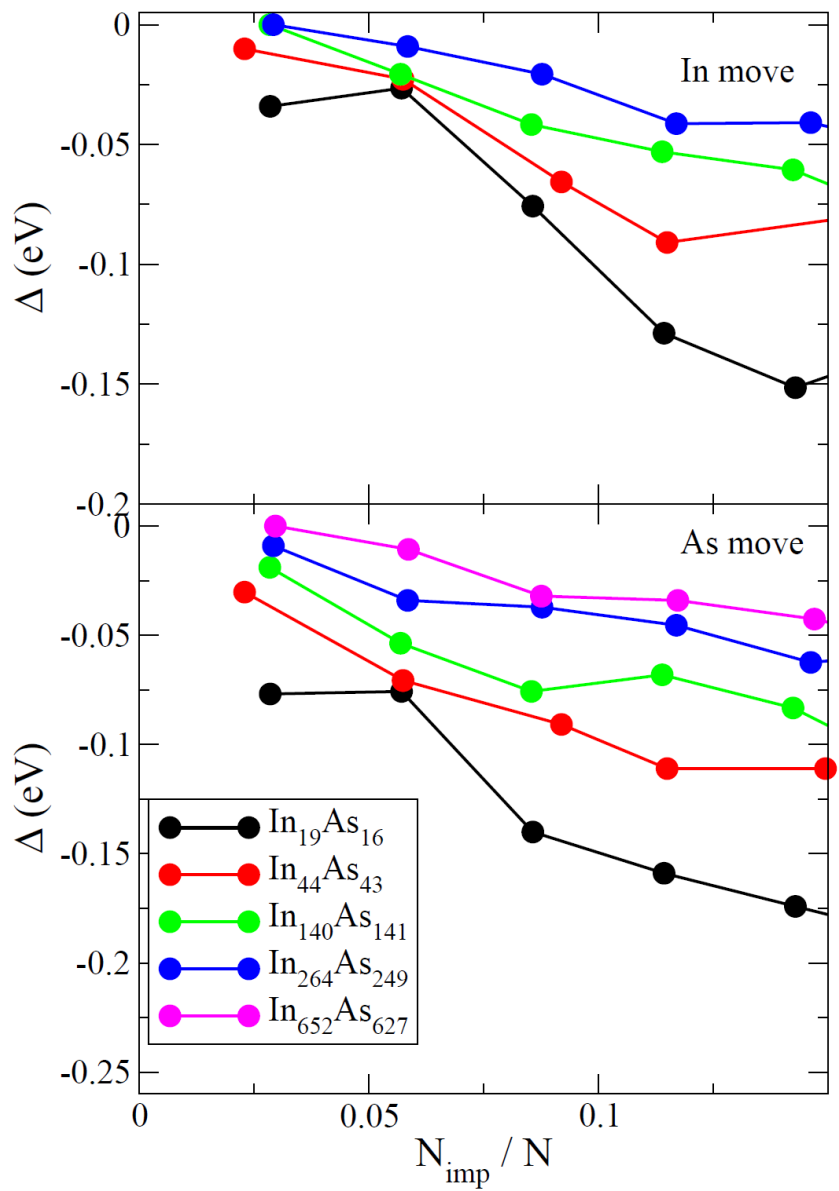

Figure S24. Energy shifts due to disorder. The average shifts (absolute value) in the DOS near the top of the valance band are plotted as a function of the ratio of the number of impurities to the total number of atoms in the nanocrystal for different nanocrystals. Upper and lower panels are for In and As disorder, respectively.